\documentclass[twoside,11pt]{article}

\usepackage{blindtext}

\usepackage[abbrvbib,preprint]{jmlr2e}

\usepackage{lastpage}
\jmlrheading{}{}{}{}{}{}{}

\ShortHeadings{Cluster MDS embedding tool for data visualization}{} 
\firstpageno{1}

\usepackage{amsmath}
\usepackage{amsfonts}
\usepackage{bm}
\usepackage{bbm}
\usepackage[normalem]{ulem}

\usepackage{comment}
\usepackage{xcolor}
\usepackage{listings}
\usepackage{color}
\usepackage{hyperref}
\hypersetup{ hidelinks }
\usepackage{fancyvrb}

\newcommand\R{\mathbb{R}}

\newcommand\ncl{N_\text{cl}}
\newcommand\nclk{N_{\mathcal{C}_k}}
\newcommand{\nlev}[1]{N_\text{level #1}}
\newcommand{\sect}[1]{Section~\ref{#1}}
\newcommand{\fig}[1]{Fig.~\ref{#1}}
\newcommand{\eq}[1]{Eq.~(\ref{#1})}

\begin{document}
\title{Cluster-based multidimensional scaling embedding tool for data visualization}
\author{\name Patricia Hern\'andez-Le\'on \email patricia.hernandezleon@aalto.fi \\
\addr Department of Chemistry and Materials Science \\
Aalto University \\
02150, Espoo, Finland
\AND
\name Miguel A. Caro \email mcaroba@gmail.com \\
\addr Department of Chemistry and Materials Science \\
Aalto University \\
02150, Espoo, Finland}

\editor{}
\maketitle

\begin{abstract}
We present a new technique for visualizing high-dimensional data called cluster MDS (cl-MDS), 
which addresses a common difficulty of dimensionality reduction methods: preserving both local and
global structures of the original sample in a single 2-dimensional visualization. Its algorithm
combines the well-known multidimensional scaling (MDS) tool with the $k$-medoids data clustering
technique, and enables hierarchical embedding, sparsification and estimation of 2-dimensional
coordinates for additional points. While cl-MDS is a generally applicable tool, we also include specific recipes
for atomic structure applications. We apply this method to non-linear data of increasing
complexity where different layers of locality are relevant, showing a clear improvement in their
retrieval and visualization quality. 
\end{abstract}


\section{Introduction}\label{s_intro}

Data complexity is a reflection of the world's complexity. This manifests itself in the
presence of high-dimensional datasets in all fields of science and the humanities.
Even though there are several types of data complexity, dimensionality alone can \textit{per
se} drastically reduce the insight (even the scientific knowledge) that we can extract from
a given sample. In this context, data visualization can be very valuable, albeit extremely
difficult to achieve with a high number of dimensions $n$ (where high means $n > 3$). 
An obvious approach to tackle this problem is reducing the number of dimensions involved, so as
to unravel the original information within our limited ``visual/dimensional grasp''.
In practice, this is essentially equivalent to finding a (satisfactory) map between a
low-dimensional
representation of the data (preferably within a 2--3 dimensional Euclidean space) and the
original high-dimensional representation.

The so-called \textit{dimensionality reduction techniques} are an example of how
to achieve one such map, increasingly used thanks to the popularization of machine learning (ML)
and data mining in most fields of science.
These methods generate an (either linear or non-linear) embedding that leads to an optimized
candidate low-dimensional representation of the original sample. Note that we refer to
\textit{candidate} representations; despite its existence, the map is not necessarily
unique nor exact~\citep{lui2018_dimloss}. There is an unavoidable tradeoff between the
amount of information
preserved and the dimensionality reduction required, leading to a plethora of possible
approaches and an extensive literature on the matter. Some of the best established among
these techniques are principal component analysis
(PCA)~\citep{pca_hotelling}, t-distributed stochastic neighbor embedding
(t-SNE)~\citep{tsne_maaten1, tsne_maaten2}, Isomap~\citep{tenenbaum_isomap} and multidimensional
scaling (MDS)~\citep{mds_kruskal1, mds_kruskal2, modern_mds}. Such variety of methods is,
however, a symptom of a deeper problem: while they aim for the same mathematical object,
their outputs differ widely. More importantly, the differing representations/visualizations
can lead to disagreement in the interpretation and retained knowledge that can be obtained
from a dataset.

Ideally, a suitable goodness-of-fit test would allow the user to choose the best algorithm
for a given sample. Unfortunately, there is no universal metric that 
measures the accuracy of the resulting embedding nor its quality as a visualization of
the original sample \citep{lui2018_dimloss, bertini2011_qualitymetrics,
tsai2012_visualmetric}, not to mention both simultaneously. Having no absolute measure
of the quality of a method, a sensible approach is to reexamine the aforementioned
techniques in order to overcome some of their particular shortcomings, usually
with the objective to either obtain an overall mathematical improvement or a
concrete domain-specific one.
Examples of the former, such as the uniform manifold approximation and projection (UMAP)
method~\citep{mcinnes2020umap}, focus on increasing the robustness of previous theoretical
foundations, with special emphasis on their mathematically based algorithmic decisions. On
the other hand, improvements for domain-specific applications are motivated by
the poor performance and visualizations obtained for high-dimensional
non-linear data, e.g., biological data in the case of the potential of heat diffusion for
affinity-based transition embedding (PHATE) tool~\citep{phate_moon}.

This paper introduces a new member of the last category that we call cluster MDS
(cl-MDS)~\citep{clmds}, motivated by previous work with atomic structures. This method arises
from the need for an embedding tool which hierarchically preserves global and local
features in
the same visualization, given their significance when analyzing atomic databases. Due
to the inherent local or global nature of the algorithms, just retaining local structures
without detriment to global ones poses a serious challenge to the majority of dimensionality
reduction techniques, especially to those favoring local distances over other scales (e.g.,
t-SNE and UMAP). In this context, methods that focus on preserving the global distance
structure (e.g., MDS and Isomap) tend to perform better, but are heavily reliant on the
distribution of local features of the dataset. 
cl-MDS is our attempt at combining the strengths of (metric) MDS with a data clustering technique. While similar ideas have already been explored
in different local MDS methods \citep{yu_HMDS_networks, shon_CMDS_networks, saeed_CBMDS_networks}, 
we use a carefully devised algorithm that always includes at least one global MDS embedding. This
is a crucial difference, that allows us to retrieve and to embed several layers of locality consistently.

Despite our background and motivation, we have developed this algorithm bearing a general approach
in mind; while the nuances of each sample may differ across domains, some of them are 
likely to possess global and local structures that could benefit from this new tool. 
As a result, this paper is organized as follows. \sect{s_algorithm} presents a detailed
description of the cl-MDS algorithm and its general features. \sect{s_examples} includes
a diverse selection of examples as well as some comparisons with other methods. The advantages and
disadvantages of cl-MDS are discussed there too. We expand on our motivation in
\sect{ss_atomic}, where we highlight the value of cl-MDS for visualizing kernel similarities of atomic
environments. We summarize and conclude in \sect{s_conclusion}. Domain-specific recipes
for atomic structure visualization are given in Appendix \ref{app_atomic}.

\section{Cluster MDS algorithm} \label{s_algorithm}

The main purpose of cl-MDS is to obtain a low-dimensional representation of some
high-dimensional data, where the distances between data points resemble the original
ones as much as possible, similarly to  metric MDS~\citep{modern_mds}. However, the key additional
constraint relates to preserving
the maximum amount of local information while improving the visualization of global
structures with a sole embedding. As discussed earlier, most of
the current dimensionality reduction methods typically fail to capture that interplay between
the local and global details, since they usually focus on either the former or the latter.

With this in mind, let us consider a sample of $N$ data points
$X = \{x_1, \hdots,\, x_N\}$ contained in a high-dimensional metric space $\R^n$, where
$n > 2$ and $\mathcal{I} = \{1,\hdots,\,N\}$ denotes its set of indices. Given an
associated distance matrix $\mathbf{D}$ with elements $D_{ij}$ for each $i,j \in \mathcal{I}$,
the intended output of the algorithm is a low-dimensional representation
$Y = \{y_1,\, \hdots,\, y_N\}$ of $X$ within $\R^2$. Unlike other methods, the
target number of dimensions $m$ is fixed ($m = 2$) because cl-MDS has been developed as a
visualization tool, rather than a general dimensionality reduction technique. Also, note that
the notion of distance refers here to any metric function defined on $X$, which will be used
as a dissimilarity measure. 

\begin{figure}[t]
\centering
\includegraphics[width=0.9\textwidth]{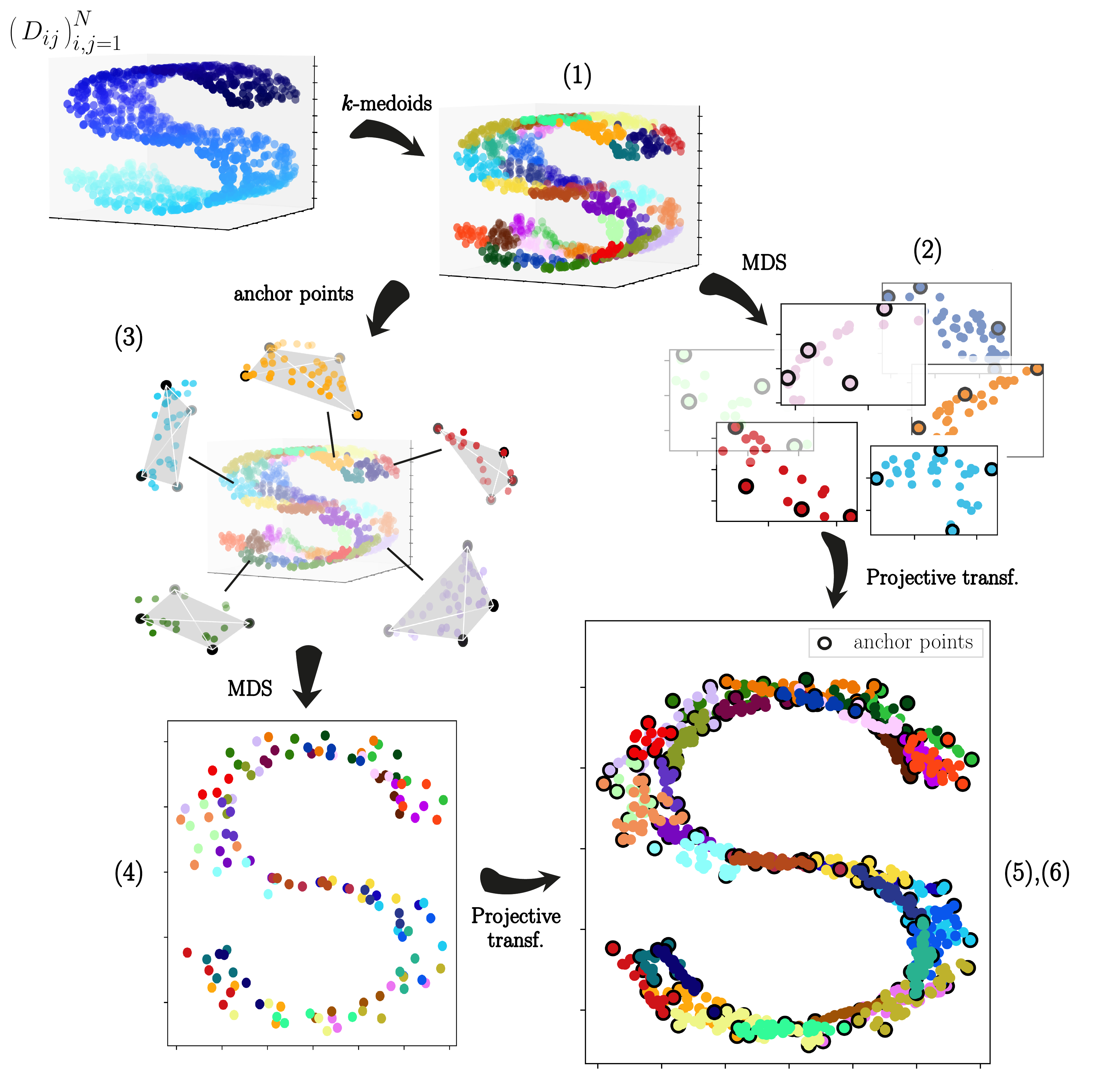}
\caption{Steps of the cl-MDS algorithm: (1) $k$-medoids clustering of the data; (2) MDS-based local
embedding of the individual clusters; (3) anchor-point selection within the individual clusters;
(4) MDS-based global embedding of the anchor points only; (5,6) global embedding of all data points
based on transformations derived from (2,4).} \label{fig_algorithm}
\end{figure}

The algorithm consists of three parts. The first and second are responsible for
identifying the local and global structure of $X$, respectively, and computing their
corresponding (independent) 2-dimensional embeddings. Since those mappings give rise to different
representations in $\R^2$, the third part of the algorithm seeks to
reconcile the information (local and global) into a single representation, leading to $Y\subset \R^2$.
More specifically, the following steps are performed (see 
\fig{fig_algorithm}):

\begin{itemize}
    \item[A.] Identify local structures
    \begin{enumerate}
        \item \label{step:1} Clustering using $k$-medoids, with $k = \ncl$, separates
        the dataset into $\ncl$ data clusters. 
        \item MDS embedding of each cluster, separately ($\ncl$ independent local maps). \label{step:2}
    \end{enumerate}
    \item[B.] Identify the global structure
    \begin{enumerate}
    \setcounter{enumi}{2}
        \item Selection of reference (\textit{anchor}) points in each cluster. \label{step:3}
        \item Joint MDS embedding of reference points only (global map). \label{step:4}
    \end{enumerate}
    \item[C.] From local to global embedding
    \begin{enumerate}
        \setcounter{enumi}{4}
        \item Checking for pathological configurations. \label{step:5}
        \item Carry local representation over to the global map, preserving the local
        structure of the data. \label{step:6}
    \end{enumerate}
\end{itemize}

These steps constitute the core of the cl-MDS algorithm and are further explained in the
following subsections.

Since the division between local and global data structure is necessarily dataset-specific,
we have extended the base cl-MDS algorithm outlined above to accommodate an arbitrarily
complex nested hierarchy of the
data structure. That is, the cl-MDS algorithm can perform an arbitrary number of levels of
embedding, in practice usually limited to a few,
by hierarchically grouping small clusters together into bigger ones. This feature can be useful for
particularly complex datasets and is introduced in \sect{ss_hierarchy}.
Additional features of the algorithm are presented in Appendix
\ref{app_atomic}, focused on a case application to representing atomic structures. 

\subsection{Local structures} \label{ss_local}

The first part of cl-MDS seeks to identify all the local information present in the sample
$X$. While there are global dimensionality reduction techniques, such as MDS, that in
principle preserve most local structures, their performance significantly deteriorates
as the size and complexity of $X$ increases (see \sect{s_examples} for examples). We
can ameliorate this problem by providing a sensible division of the sample into subsets
of related data points and then computing a 2-dimensional embedding individually for each
subset. Though it might seem that we are replacing one problem with another, this solution
is quite convenient for our purposes as will become clear in the next subsections.
Additionally, since the algorithm and its output are based on dissimilarity data, we have
a natural criterion for a classification into subsets. Using that information, we can
divide $X$ in $\ncl$ subsets known as \textit{clusters}. Their main characteristic is the
low dissimilarity values between members of a same cluster compared with those for points in
different ones. As a result, each subset has a smaller distribution of dissimilarities than
the complete sample, which will also help in our search for a finer embedding. 

Therefore, the cl-MDS algorithm starts with the computation and optimization of a clustering of $X$
(step \ref{step:1}), with the distance matrix $\mathbf{D}$ and the total number of clusters 
$\ncl$ as input parameters. A suitable clustering technique is the $k$-medoids
method~\citep{kaufman_kmedoids, bauckhage_kmedoids}, with $k = \ncl$ in our case.
Unlike similar (and arguably more popular) barycenter-based algorithms such
as $k$-means \citep{macqueen_kmeans,hartigan_kmeans} or spectral clustering \citep{Ng_spectral_clustering},
$k$-medoids builds each cluster considering a \textit{medoid}. This is the element in a cluster with
minimal average dissimilarity to the remaining points in that same cluster. Thus, the $k$-medoids
clustering process relies on selecting actual points from the sample as centroids, rather than
these centroids being calculated as the coordinates in $\R^n$ with, respectively, the smallest
average intracluster distances. The distinction
is important because, even though all points in $X$ belong to $\R^n$, not all points in $\R^n$
can necessarily be mapped back to a meaningful or interpretable data point. We provide
a concrete example of this lack of bijection in Appendix \ref{app_atomic} for the
smooth overlap of atomic positions (SOAP) high-dimensional representation of atomic
structures~\citep{bartok_soap}, which motivates our choice of clustering technique.

For computational efficiency, we have reimplemented Baukhage's $k$-medoids 
Python recipe~\citep{bauckhage_kmedoids} in Fortran. Our implementation, that we call
\texttt{fast-kmedoids}~\citep{fast_kmedoids}, can be easily built into a Python package
with \texttt{F2PY}~\citep{peterson_2009}. \texttt{fast-kmedoids} incorporates several other
new features beyond increased speed, including optimization of initial medoid selection. The latter
constitutes one of the main problems with centroid-based techniques, whose result heavily relies on
the initialization. Since random selection is too volatile on its own, we combine it with
\textit{farthest point sampling} to select the \verb|n_iso| most isolated points from the sample. 
By default, we set \verb|n_iso|~$=1$. The complete clustering process in cl-MDS is performed as follows:
    \begin{itemize}
        \item[1.1] Use the $k$-medoids algorithm to obtain a set of clusters  $\mathcal{C}=\{\mathcal{C}_1,
        \hdots ,\,\mathcal{C}_{\ncl} \}$ and their corresponding medoids $\mathcal{M}=\{m_1,\hdots 
        ,\, m_{\ncl}\}$, where $m_k\in \mathcal{C}_k$, $\mathcal{C}_k \subset \mathcal{I}$ and 
        $x_{m_k}\in X$ for any $k=1,\hdots ,\ncl$. 
        \item[1.2] Compute the relative intra-cluster \textit{incoherence}~\citep{caro_2018_aC},
        \begin{align}
        \text{I}_\text{rel} = \sum_{k=1}^{\ncl} \frac{1}{\nclk} \left(\sum_{i\in \mathcal{C}_k} D_{i,m_k}\right)\,,
        \end{align}
        where $\nclk$ denotes the cardinality of $\mathcal{C}_k$.
        \item[1.3] Repeat until a fixed maximum number of tries is reached,
        for a different medoid initialization. Keep the set with minimal
        $\text{I}_\text{rel}$, ensuring the lowest internal incoherence of the clusters among all
        candidate $k$-medoids solutions.
    \end{itemize}
In our cl-MDS Python implementation~\citep{clmds}, the number of repetitions is set by the parameter
\verb|iter_med|, while the initialization of
$k$-medoids is controlled by \texttt{init\_medoids = "random" | "isolated" (default)}.
The default option forces the inclusion of (at least) the most isolated point in the initialization.
We recommend to test different proportions of isolated initial medoids (\texttt{n\_iso\_med}), especially
for highly uneven data distributions where certain data types are infrequent but very relevant (e.g., when
interesting areas of the configuration space are poorly sampled). Also, note that extensive iteration is
needed for a robust initialization, and it is not guaranteed to find the global minimum configuration of
medoids. The user can provide their custom initial medoids, e.g., when opting for a more thorough search
using alternative optimization approaches.

Next, we proceed with the embedding, step \ref{step:2}. A low-dimensional 
representation $Y_{\mathcal{C}_k}^{(l)}$ for each cluster $\mathcal{C}_k$ is computed applying
the standard MDS method, with $k=1,\hdots ,\ncl$. In particular, we use a weighted variation
\citep{clmds} of the metric MDS implementation included in the Python module 
\verb|scikit-learn|~\citep{sklearn}. In this version of the method, the coordinates in the low-dimensional
space are optimized such that the pairwise distances between the embedded data points
reproduce the input dissimilarity data as closely as possible. This is achieved by minimizing the
stress, an objective function defined as 
\begin{align} \label{eq:str}
    \sigma = \sum_i \sum_{i>j}  w_{ij}\left(D_{ij} - d_{ij}\right)^{2}\,,
\end{align}
where $d_{ij}$ and $w_{ij}$ correspond to the computed distances in the low-dimensional space and
their weights respectively, with $i,j \in \mathcal{I}$. Therefore, each data point gets an optimized
representation using a non-linear map. Note that we need to partition $\mathbf{D}$ considering the
previous set of clusters before computing the MDS embeddings; that is, each representation
$Y_{\mathcal{C}_k}^{(l)}$ is obtained from a stress $\sigma_k$ restricted to $i,j \in \mathcal{C}_k$.
An additional remark: the current cl-MDS implementation does not address missing values in the distance matrix, since it only accepts weights per cluster ($w_k$ for all $i,j \in \mathcal{C}_k$). That said, the code could be carefully adapted in the future to accommodate missing dissimilarities~\citep{groenen_2016_mds_review}; meanwhile, a data-specific pre-processing step is needed in such case. Similarly, we could weight the importance of each data point in the stress inversely to some expected noise, handling noisy data.

Once step \ref{step:2} concludes, a set of embeddings $Y^{\text{(local)}}=\{Y_{\mathcal{C}_1}^{(l)},
\hdots,\,Y_{\mathcal{C}_{\ncl}}^{(l)}\}$ is obtained. Hence, the output is a collection of $\ncl$
2-dimensional representations containing distinct ``slices'' of local information, despite reaching
the same target space. As we will show next, they do not correspond to the final representation
$Y$ of our algorithm either.

\subsection{Global structures}\label{ss_global}

The second part of the cl-MDS algorithm gathers the global information and maps it to a
2-dimensional space $\R^2$. As for the local embeddings, we utilize the
same dimensionality reduction technique, MDS. As we discussed in \sect{ss_local}, MDS
mapping of the whole sample may lead to unsatisfactory results for large datasets, whereas applying it
to a small subset made of selected points could give us enough insight into its global features.
That subset must also contain several reference points from each cluster, enclosing as much local
information as possible. Since the distribution of those points also determines the global map, they
will act as \textit{anchor points} between local and global representations. Therefore, this part
of cl-MDS is divided in two steps: finding a suitable collection of reference points that ``frames''
all clusters (step \ref{step:3}) and, then, proceeding with their embedding (step \ref{step:4}).

Step \ref{step:3} selects as anchor points a subset of $X$ aiming at satisfying
two conditions simultaneously: preserving a maximal amount of local information and outlining the
overall sample features. Intuitively, we need at least one anchor point from each cluster to grasp
their global distribution. However, this would disregard the local structure within the clusters.
Therefore, in practice, up to four anchor points per cluster ($1\leq n_\text{anc}\leq 4$)
are chosen to both ensure the aforementioned preservation of local features and ease the MDS
minimization problem.
The benefits of this choice are further discussed on \sect{ss_localtoglobal}. Hence,
this part of the algorithm searches the high-dimensional representation of each cluster $\mathcal{C}_k$
for the vertices $\mathcal{A}_k =\{a_{k1}, \hdots, a_{kn_{\text{anc}}}\} \subset \mathcal{C}_k$ which
define the tetrahedron of maximal volume, for $k=1,\hdots,\,\ncl$. The detailed procedure is as follows:
\begin{itemize}
    \item[3.1] Check if the cardinality $\nclk$ of cluster $\mathcal{C}_k$ satisfies 
    $\nclk>\nclk^\text{max}$. For
    the sake of computational efficiency, large clusters benefit from discriminating between points to reduce
    the list of candidate vertices before considering any tetrahedra. In that case, the $p$-th percentile
    of the distance to the medoid $m_k$ is chosen as threshold, leaving only farther points in
    $\mathcal{C}_k$ as vertices. The percentile rank $p$ is customized for different $\nclk$ with the
    parameter \verb|param_anchor|. Our tests indicate that $\nclk^\text{max} = 70$ is a robust choice, and
    this is hardcoded in the cl-MDS program.
    \item[3.2] Compute the volume of each possible tetrahedron whose vertices $v$ are a subset of the 
    (reduced) list of points in $\mathcal{C}_k$. Since a tetrahedron corresponds to a 3-simplex $S_3$,
    its volume $V$ can be computed using the Cayley–Menger determinant as follows 
    \citep{Sommerville_geomNdim, gritzmann_klee_1994}:
    \begin{align}
        V(S_3)^2 = \frac{1}{2^3 (3!)^2}\, 
        \begin{vmatrix}
            0 & \mathbf{1}_4^\top \\
            \mathbf{1}_4  & \mathbf{D}[v] 
        \end{vmatrix}\,,
    \end{align}
    where $\mathbf{1}^\top_4 =(1,1,1,1)$ and $\mathbf{D}[v]$ is the $4\times 4$ submatrix of $\mathbf{D}$
    corresponding to the vertices of the 3-simplex. We note here that the tetrahedron's volume is computed
    directly from the distances, and that these distances are not necessarily based on an Euclidean metric.
    \item[3.3] Keep the set of vertices $\mathcal{A}_k$ forming the tetrahedron with maximal volume.
\end{itemize}

Repeating those steps for each cluster, we obtain the complete set of anchor points 
$\mathcal{A} = \bigcup_{k=1}^{\ncl}\mathcal{A}_k$. We opted for the volume as a measure of the amount
of preserved information, although one could argue that there exist better criteria, such as the
number of data points enclosed by the tetrahedron. Those measures are nonetheless much more expensive
computationally given their dependence on $\nclk$, whereas the volume criterion is independent of it.

Finally, step \ref{step:4} is carried out. Applying the same weighted metric variation of the MDS
method from step \ref{step:2}, a 2-dimensional representation of the anchor points 
$Y^{\text{(g)}}_{\text{anchor}} = \{y_{a_i}^{\text{(g)}}\,|\, a_i\in \mathcal{A}\}$ is computed.
Notably, those new coordinates encode the global 2-dimensional distribution of the local structures
previously clustered.

\subsection{From local to global embedding }\label{ss_localtoglobal}

We have so far obtained a \textit{collection} of low-dimensional representations of the original dataset
$X$, characterized by preserving local and global structures \textit{separately}. These embeddings
are limited to certain subsets of points, i.e., anchor points in the case of the global one and clusters
in the case of the local ones. The last part of the algorithm seeks to reconcile these two independent
representations to achieve a complete representation $Y$ of $X$, by mapping each cluster's local coordinates
into the global map. That is, given a cluster $\mathcal{C}_k$ with $k=1,\hdots,\,\ncl$, we can leverage
its anchor points $\mathcal{A}_k$, whose coordinates are known in local and global representations, to
obtain a suitable transformation for all of its members. 

Two types of transformations are implemented: affine transformations $\mathbf{A}$,
and projective transformations $\mathbf{H}$ (also known as perspective mappings or \textit{homographies})
~\citep{schneider_geometric_tools}. The simpler affine transformations
preserve parallel lines, as opposed to the non-linear maps performed by the MDS algorithm. Hence, homographies
were introduced for the sake of consistency, since they always preserve incidence (relations of containment
between points, lines and planes) but not parallelism. There are other properties that are not preserved by
the MDS algorithm, suggesting that there is room for future improvement in this direction. For instance, the
relative ordering of distances between points is not always preserved, despite the MDS efforts of reproducing
relative distances
[see Eq.~(\ref{eq:str})]. This property will influence the selection process in step \ref{step:5}, as we explain
below.

Before computing a transformation operator (in the form of a matrix),
we need to ensure the absence of pathological configurations.
Step \ref{step:5} determines the most suitable transformation (affine or projective) for each cluster, depending
on the number of anchor points $n_{\text{anc}}$ and their relative distribution in both representations. The
importance of this last property lies in its relation to convexity, which is a necessary (and sufficient)
condition for homography. That is, the anchor points $\mathcal{A}_k$ ($k=1,\hdots,\,\ncl$) form a quadrilateral
in each representation whose convexity is not ensured in both maps (as anticipated above), potentially leading
to an ill-conditioned transformation. Thus, the pathology check for a cluster $\mathcal{C}_k$ is performed as
follows:
\begin{itemize}
    \item[5.1] Check if $n_{\text{anc}}<4$. In that case, an affine transformation $\mathbf{A_k}$ is chosen
    trivially and no further steps are needed.
    \item[5.2] Check if $\mathcal{A}_k$ does not coincide with its convex hull (i.e., the smallest convex subset
    in $\mathcal{A}_k$ containing $\mathcal{A}_k$) in the local representation $Y^{(l)}_{\mathcal{A}_k}$.
    This corresponds to one of the anchor points being enclosed within the triangle defined by the
    other three, or to (at least) three anchor points being collinear. In that
    case, a homography is ill-conditioned and we redefine $\mathcal{A}_k$ as its convex hull, implying
    $n_{\text{anc}}<4$ and obtaining the result of step 5.1. The convex hull is computed using the Qhull
    library~\citep{qhull} through \verb|scipy.spatial.ConvexHull|~\citep{2020SciPy-NMeth}.
    \item[5.3] Repeat the previous check, now using the global representation $Y^{(g)}_{\mathcal{A}_k}$. If true,
    the four anchor points will be used to compute an affine transformation $\mathbf{A_k}$. Otherwise, the convexity
    is confirmed in both representations and a homography $\mathbf{H_k}$ is chosen.
\end{itemize}

Once each cluster has been checked, the algorithm can proceed with their mapping using the most appropriate
transformation $\mathbf{T_k}$ for each $k=1,\hdots,\,\ncl$, where $\mathbf{T_k}$ corresponds to $\mathbf{A_k}$ or
$\mathbf{H_k}$. For simplicity, we use homogeneous coordinates to express the details of step \ref{step:6}. In this
notation, any point $p^\top=(x,y)$ from $\mathbb{R}^2$ has an homogeneous representation $\Tilde{p}^\top=(x,y, 1)$.
On the other hand, any homogeneous point $\Tilde{q}^\top=(u,v,w)$ is identified with 
$\Tilde{r}^\top=\Tilde{q}^\top/w=(u/w,v/w,1)$, implying the existence of a bijection between Cartesian coordinates
in the Euclidean plane and homogeneous coordinates in the \textit{projective plane} ($\mathbb{P}^2$). Alternatively, 
equivalence classes can be used to understand the concept of homogeneous coordinates within a more rigorous theoretical
framework~\citep{projective_geom}. This notation
allows us to represent the transformations as simple matrix multiplications, whose details are explained below.
Then, step \ref{step:6} is organized as follows:

    \begin{itemize}
        \item[6.1] Compute the transformation matrix $\mathbf{T_k}$ that solves the equation
            \begin{align} \label{eq_transf}
    \begin{pmatrix}
         y^{\text{(g)}}_{a_{i}}\\
         1
   \end{pmatrix} =  \mathbf{T_k} \begin{pmatrix}
         y^{\text{(l)}}_{a_{i}}\\
         1
   \end{pmatrix} \,,
    \end{align}
    for each $a_i\in\mathcal{A}_k$. 
    
    In particular, affine transformations (usually represented as the composition of a linear transformation
    $\mathbf{L_k}$ and a translation $\mathbf{b_k}$) correspond to the following matrix,
\begin{align}
    \mathbf{A_k} = \begin{pmatrix}
           \mathbf{L_k} & \mathbf{b_k}\\
           \mathbf{0}_n^\top &  1
        \end{pmatrix}\,,
\end{align}
where $\mathbf{0}_2^\top=(0,0)$ for transformations in $\mathbb{R}^2$, and, therefore, they are obtained as the
least-squares solution of Eq.~(\ref{eq_transf}). 

On the other hand, a homography matrix~\citep{eberly_perspective_map} has the form 
\begin{align}
        \mathbf{H_k} = \left(\mathbf{A_k^{(g)}}\right)^{-1}\mathbf{F} \;\mathbf{A_k^{(l)}}\,,
    \end{align}
    where $\mathbf{A_k^{(l)}}$ is the affine transformation from the local $\mathcal{A}_k$ coordinates to the
    \textit{canonical quadrilateral} $\{(1,0),(0,0),(0,1),(a,b)\}$, with $(a,b) = \mathbf{A_k^{(l)}} y_{a_3}^{\text{(l)}}$.
    Likewise, $\mathbf{A_k^{(g)}}$ is the equivalent transformation from the global coordinates to a canonical
    quadrilateral whose fourth point is $(c,d) = \mathbf{A_k^{(g)}} y_{a_3}^{\text{(g)}}$. Both affine transformations
    can be computed using the least-squares method too. Finally, $\mathbf{F}$ is a \textit{linear fractional
    transformation} from the first canonical quadrilateral into the second,
    \begin{align}
        \mathbf{F} = \begin{pmatrix}
            bcs & 0 & 0 \\
            0 & ads & 0 \\
            b\,(cs-at) & a\,(ds-bt) & abt \\
        \end{pmatrix}\,,
    \end{align}
    where $s=a+b-1$ and $t=c+d-1$ are positive for convex quadrilaterals. Since both pairs $(a,b)$ and $(c,d)$ are
    known from the previous affine transformations, $\mathbf{F}$ is known too.

        \item[6.2] Compute the global representation of cluster $\mathcal{C}_k$ using the previous mapping,
    \begin{align}
    \begin{pmatrix}
         y'_{i}\\
         w
   \end{pmatrix} =  \mathbf{T_k} \begin{pmatrix}
         y^{\text{(l)}}_{i}\\
         1
   \end{pmatrix}\,,
    \end{align}
    and applying the \textit{perspective divide} $y_i^{(g)} = y'_i/w$ to each $i\in\mathcal{C}_k$.
    \item[6.3] If $\mathbf{T_k}=\mathbf{H_k}$, an affine transformation is also computed for comparison. We retain
    the result with the lowest residue,
    \begin{align}
    R(\mathbf{T_k}) = \sum_i(y^{(l)}_i-y_i^{(g)}(\mathbf{T_k}))^2    \,,
    \end{align}
    in order to minimize the effect of outliers.
    \end{itemize}

As a result of step \ref{step:6}, we obtain the output of the cl-MDS algorithm,
$Y=\bigcup_{k=1}^{\ncl}Y^{\text{(g)}}_{\mathcal{C}_k} =\{y_1,\hdots,\,y_N\}$.
These steps constitute the core of the cl-MDS algorithm.

\subsection{Cluster hierarchy and sparsification} \label{ss_hierarchy}

\begin{figure}[t] 
\centering
\fbox{
\includegraphics[width=0.9\columnwidth,trim={0.5cm 0 0.5cm 0.1cm},clip]{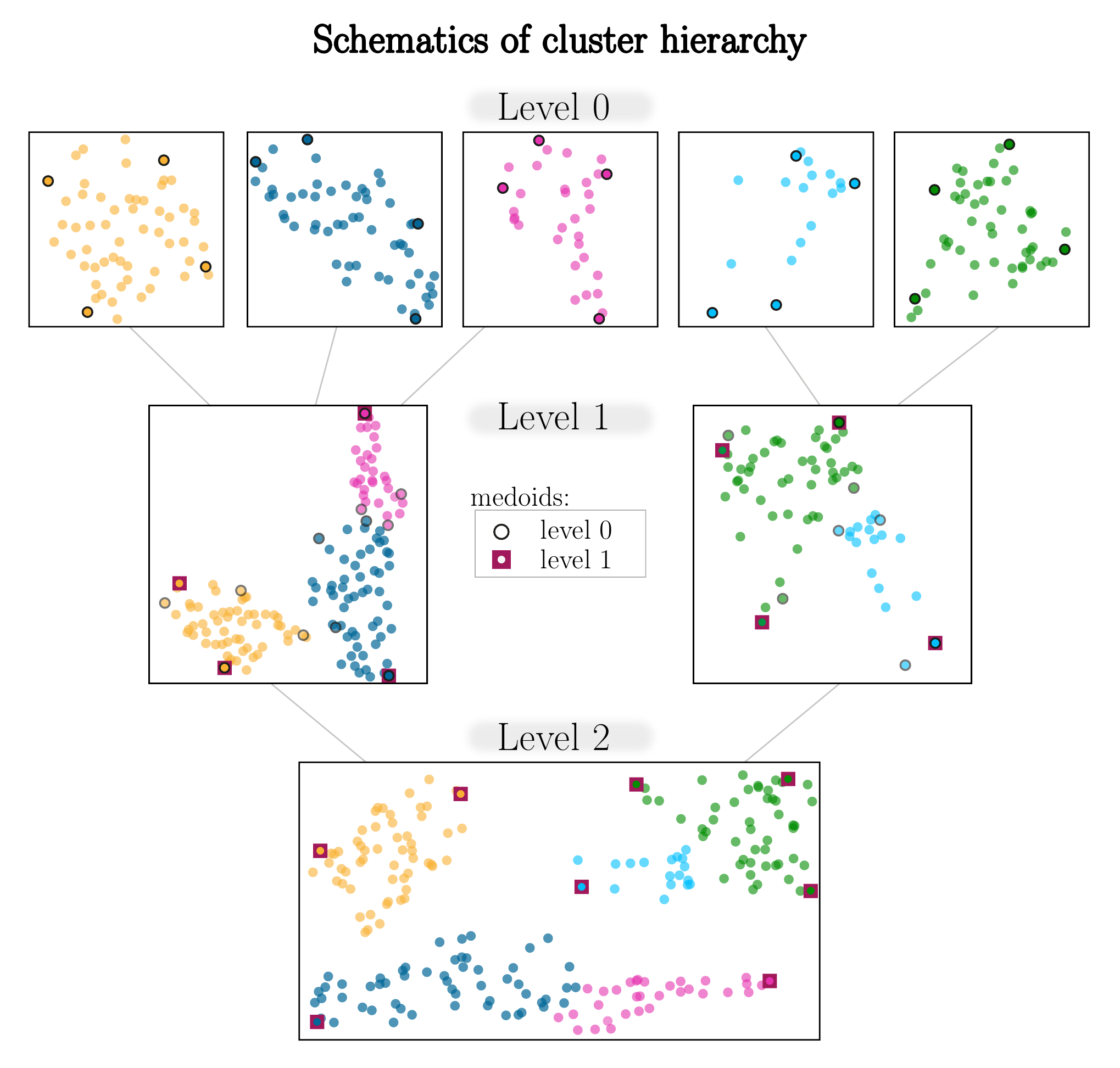}
}
\caption{Illustration of the hierarchical cluster setup, using a $[5,2,1]$ hierarchy. The figure includes
the set of anchor points obtained per clustering level, following step~\ref{step:3}. The last level does
not require such set since it corresponds to the final embedding.}\label{fig_hierarchy}
\end{figure}

Additional features are implemented to complement and improve the cl-MDS algorithm. Here we introduce the general ones,
whereas Appendix \ref{app_atomic} details those specific to visualization of databases of atomic structures.
While cl-MDS strives to preserve
local and global structures, its base algorithm sometimes lacks the sufficient flexibility for mapping complex databases
where different layers of locality can be meaningful. To address this issue we introduce \textit{cluster hierarchy}.
In a hierarchical cluster setup, the
number of clusters hyperparameter $\ncl$ is replaced by a hierarchy hyperparameter, 
\begin{align}
    h=[\nlev{0},\, \nlev{1},\, \nlev{2},\hdots,\,1]\,,
\end{align}
 where $\nlev{0}$ refers to the finest clustering level and $1$ represents the final global MDS embedding (i.e.,
 $\nlev{0} > \hdots > \nlev{m} > \hdots > 1$). This list enables a hierarchical embedding on 
 steps~\ref{step:3}~to~\ref{step:6} that mimics the
 idea behind hierarchical clustering \citep{ESL}. Each level $m$ corresponds to a grouping $\mathcal{C}^{\text{($m$)}}$ of
 the dataset with a specific embedding $Y^{\text{($m$)}}$ in the 2-dimensional Euclidean space, such that the next level
 ($m+1$) merges several of those clusters and computes a new 2-dimensional embedding $Y^{\text{($m+1$)}}$, using $Y^{\text{($m$)}}$
 as the local representation. In this way, hierarchical embedding improves the representation of samples with several
 levels of locality. The simplest hierarchy $[\ncl, 1]$ is equivalent to $\ncl$, where a sole
 layer of local information is considered. The hierarchy approach is schematically depicted in Fig.~\ref{fig_hierarchy}
 for a [5,2,1] hierarchy. See \sect{s_examples} for examples.
 
To enable processing of large databases we have added sparsification support to cl-MDS. In our Python implementation
this is selected through the keywords \verb|sparsify| and \verb|n_sparse|. Sparsification carries out the cl-MDS
operations only on a subset of the data points, selected according to some predefined recipe
(\texttt{sparsify = "random" | "cur" | list}). \texttt{n\_sparse} determines the (maximum) size of the resulting
sparse set, denoted $\mathcal{I}_{sp}$. Three sparsification options are implemented. The first selection method, 
\texttt{"random"}, uses \verb|numpy|'s random sampling
routines~\citep{numpy_2020}. The second option, \texttt{"cur"}, is based on CUR matrix
decomposition~\citep{cur_decomp, tensor_cur_decomp},
a low-rank approximation procedure. This decomposition is characterized by retaining those rows (and columns) from the
original matrix that allow its best low-rank fit, i.e., that capture the most representative part of it. The 
``significance'' of a row can be expressed in terms of its statistical influence~\citep{cur_statleverage} or its
Frobenius norm~\citep{mining_datasets}, among other properties. The CUR implementation included in cl-MDS uses the latter. 
Hence, given a distance matrix $\mathbf{D}$ and a rank \verb|n_sparse|, we choose the indices of those rows with 
higher Frobenius norm within the dataset. As a third sparsification option, the user can provide directly a custom
list or array with the sparse indices to \verb|sparsify|.

Bearing sparsification in mind, we developed a supplementary algorithm (step 7) that estimates the low-dimensional
representation of those data points not included in the sparse set, i.e.,
$\overline{\mathcal{I}}_{sp}:=\mathcal{I}-\mathcal{I}_{sp}$. The embedding transformations inferred for the sparse
set are reused for embedding $\overline{\mathcal{I}}_{sp}$, whereas the MDS mapping per cluster is replaced by an affine
mapping. This is achieved through a three-step process:
\begin{itemize}
    \item[7.1] Assign each data point $i$ in $\overline{\mathcal{I}}_{sp}$ to the same cluster $\mathcal{C}_k$ as
    its nearest medoid $m_k$, with $k\in\{1,\hdots ,\ncl\}$. This approach extends the existing clustering to the
    complete database without requiring computing or storing its complete distance matrix. To avoid future confusion,
    we refer to the extended cluster $k$ as $\mathcal{C}^*_k$, implying that 
    $\mathcal{C}_k = \mathcal{C}^*_k\cap\mathcal{I}_{sp}$.
    
    \item[7.2] Compute an affine mapping from the original high-dimensional space $\mathbb{R}^n$ to $\mathbb{R}^2$,
    for each cluster $\mathcal{C}^*_k$ with $k=1,\hdots ,\ncl$. Its matrix representation $\mathbf{\widetilde{A}}_k$
    is the least-squares solution to the equation
 \begin{align}
    \begin{pmatrix}
         y^{\text{(l)}}_{j}\\
         1
   \end{pmatrix} = \mathbf{\widetilde{A}_k} \begin{pmatrix}
         x_{j}\\
         1
   \end{pmatrix} \,,
\end{align}
for all sparse points $j\in\mathcal{C}_k$. The tilde is included to emphasize that the embedding is between distinct
spaces, as opposed to the affine transformations from step \ref{step:6}. 

    \item[7.3] Obtain the 2-dimensional representation of each cluster $\mathcal{C}^*_k$ via the composition 
    $\mathbf{\widetilde{T}_k} := \mathbf{T_k}\, \mathbf{\widetilde{A}_k}$, with $k=1,\hdots ,\ncl$. Similarly
    to step \ref{step:6}.2, these coordinates are the result of the perspective divide $y_i^{(g)}=y'_i/w$, defined
    by the equation
    \begin{align}
    \begin{pmatrix}
         y'_{i}\\
         w
   \end{pmatrix} =  \mathbf{\widetilde{T}_k} \begin{pmatrix}
         x_{i}\\
         1
   \end{pmatrix}\,,
\end{align}
where $i\in\mathcal{C}^*_k\cap\overline{\mathcal{I}}_{sp}$. Note that we exclude the data points from
$\mathcal{C}_k$ from these calculations because their 2-dimensional representation is already known.
\end{itemize}

Therefore, this algorithm completes the 2-dimensional representation of the database, $Y=\{y_1,\hdots,\,y_N\}$, previously
restricted to the sparse set, $Y_{sp}:=\{y_j\,|\,j\in \mathcal{I}_{sp}\}$. Note that the estimation will be as good (or as bad)
as the chosen sparse set.

\section{Examples} \label{s_examples}

In this section, we analyze several examples to show the potential of the cl-MDS method as well as its weaknesses.
First, a variety of toy examples are introduced in \sect{ss_toy} to illustrate its main features and to compare
its performance with other dimensionality reduction techniques. Instances of higher complexity are shown in
Section \ref{ss_atomic}, regarding several extensive atomic databases. Additional details of our motivation and these datasets are given in their respective
subsections. We decided to focus on atomic-structure samples because (i) they were our
motivation to develop cl-MDS in the first place, and (ii) we wanted to apply all the functionalities
available, including those related to atomic
structure representations. However, the cl-MDS method is applicable to datasets from other fields too.

The following algorithms are considered for qualitative comparisons with cl-MDS: locally linear embedding
(LLE)~\citep{roweis_lle}, modified LLE~\citep{zhang_mlle}, Hessian eigenmaps (Hessian LLE)~\citep{donoho_hessianlle}, local
tangent space alignment (LTSA)~\citep{zhang_ltsa}, Laplacian eigenmaps (LE)~\citep{belkin_le}, and the already mentioned
Isomap~\citep{tenenbaum_isomap}, PCA~\citep{pca_hotelling}, kPCA~\citep{kpca_1998}, 
MDS~\citep{mds_kruskal1, mds_kruskal2}, t-SNE~\citep{tsne_maaten1,tsne_maaten2} and UMAP~\citep{mcinnes2020umap}.
They are computed using \verb|scikit-learn|~\citep{sklearn} implementations except for UMAP, which has its
own Python module (\verb|umap|). On the other hand, detailed quantitative comparisons (the so-called
goodness-of-fit tests) are not included, since a good choice of metric is highly dependent on the application domain,
the user's expectations and the sample itself~\citep{bertini2011_qualitymetrics}. These metrics are (inevitably)
tailored to diverse definitions of accuracy and visual quality, further adapted to account for specific
shortcomings (e.g., the DEMaP metric~\citep{phate_moon}, the ``local continuity'' (LC) meta-criteria~\citep{chen_lmds},
or measures of \textit{trustworthiness} and \textit{continuity}~\citep{venna_lmds}). Therefore, we opted for
strong qualitative comparisons that provide a broader taste of cl-MDS, but we encourage the user to apply
these metrics to their own samples after careful selection.

As a consequence of its algorithm, cl-MDS inherits several characteristics of MDS which are relevant for
understanding the figures in this section. While their embedding dimensions lack interpretability (i.e., absence of
meaningful axes), the relative Euclidean distances of the resulting 2-dimensional representation encode the original
dissimilarity measure. Thus, the nearer two data points are in this visualization, the more similar they are and
vice versa. In practice, this does not always hold for MDS, a problem that cl-MDS alleviates as we will discuss
in \sect{ss_cho}. Additionally, the embedded representation is invariant under affine transformations for both
methods. However, the embedding codomain of cl-MDS is usually $[-1,1]\times [-1,1]$ without being strictly
restricted to it, as opposed to MDS. Since dimensionality reduction techniques differ on their codomain, as well
as on their interpretation (only PCA has a strong meaning associated to its axes), we omit the axis information in
all the figures to avoid misleading comparisons.

\subsection{Toy examples}\label{ss_toy}
Before diving into complex examples, let us discuss the main features of cl-MDS using simpler datasets. We start
following a classic \verb|scikit-learn| example of manifold learning methods~\citep{S_example}, where an S-curve
dataset with 1000 points and its corresponding Euclidean distances are used. We chose minimal
parameters for all the techniques to compare their fastest computational speed. However, note that cl-MDS was not intended
for improving time performance and has not been fully optimized accordingly. Also, this is the only example
where the parameters are not fine tuned.

\begin{figure}[t]
    \centering
    \includegraphics[width=1\textwidth]{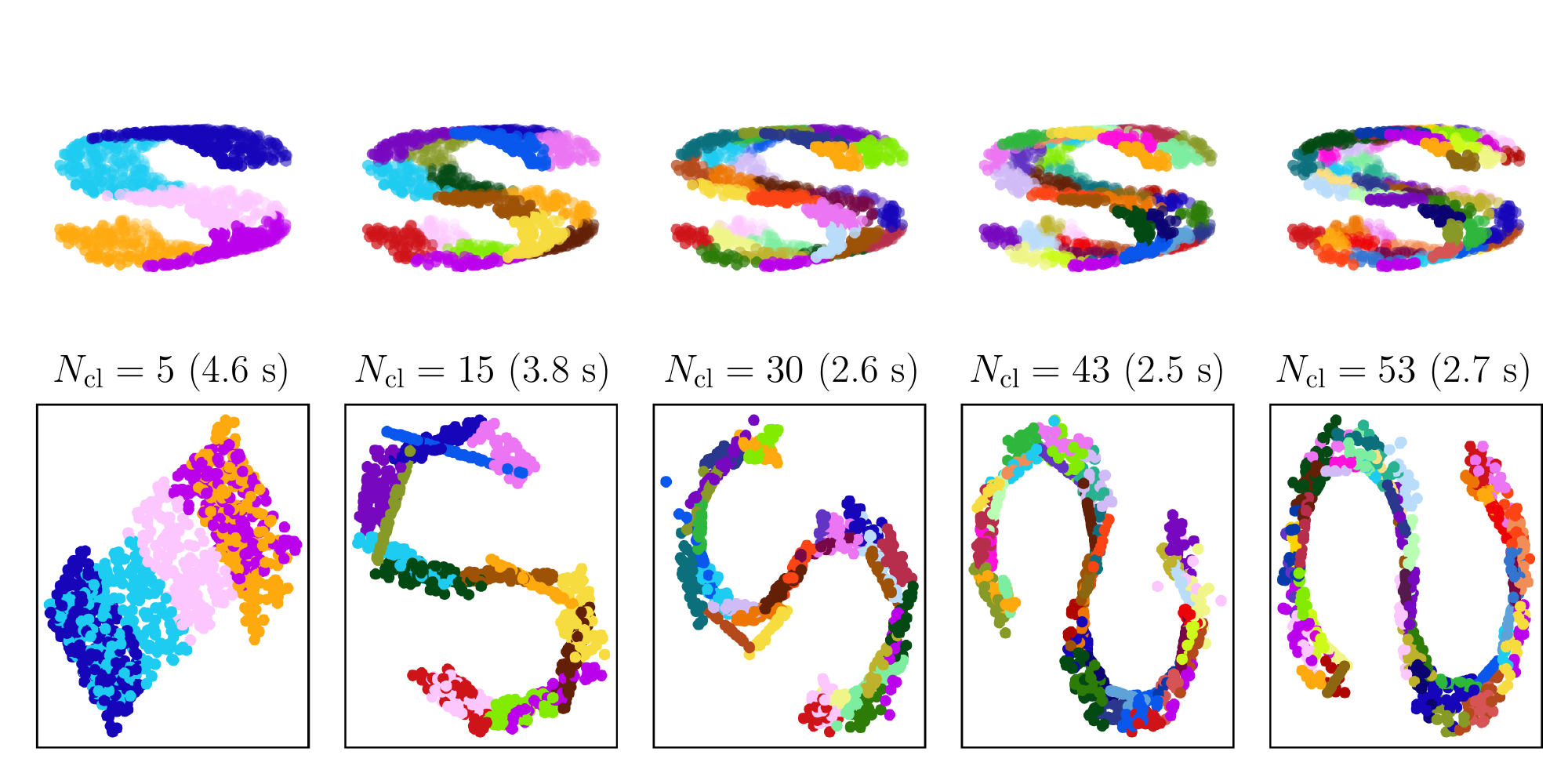}
    \caption{Effect of the number of clusters $\ncl$ in cl-MDS embedding and performance for a simple example.}
    \label{fig:n_clusters}
\end{figure}

Figure~\ref{fig:n_clusters} illustrates the effect of the cl-MDS main hyperparameter, the number of clusters $\ncl$
(see \sect{ss_local}). As expected, the algorithm output is sensitive to this choice, requiring a minimum
clustering ($\ncl = 15$) to preserve the distances in the embedded space consistently. Just like MDS, cl-MDS
focuses on revealing the metric structure, which does not guarantee uncovering the low-dimensional manifold.
This is especially true when using Euclidean distances, since they do not characterize the global distribution
of the manifold properly. Hence, this particular example requires feeding a larger amount of local information
to the global MDS mapping (step \ref{step:4}) through the clustering, in order to counteract the ambiguity in
large distances. 

In general, a good choice of $\ncl$ depends on the database and the provided metric, although a number between
5 and 20 is enough in our experience to capture local and global details properly. A higher number of clusters
can be useful to retrieve finer details, especially in highly complex datasets. Nevertheless, overly increasing
$\ncl$ worsens the global embedding and dilutes the intrinsic value of medoids as representative data points.
Therefore, we recommend finding a compromise between a finer embedding and a smaller clustering. In particular
cases, a carefully devised cluster hierarchy can prove quite helpful; the added versatility of hierarchy and
medoids information will be discussed in depth in Sections \ref{ss_cho} and \ref{ss_qm9}. Although we use a
heuristic approach for fine-tuning most of the parameters in this article, there 
are well-known criteria for choosing a suitable number of 
clusters~\citep{liu_cluster_val_measures, halkidi_cluster_validation}. We have already explored some options,
such as the elbow method and the silhouette statistic~\citep{kauffman_cluster_analysis}, but further analysis
is needed in this direction.

\begin{figure}[t]
    \centering
    \includegraphics[width=1\textwidth]{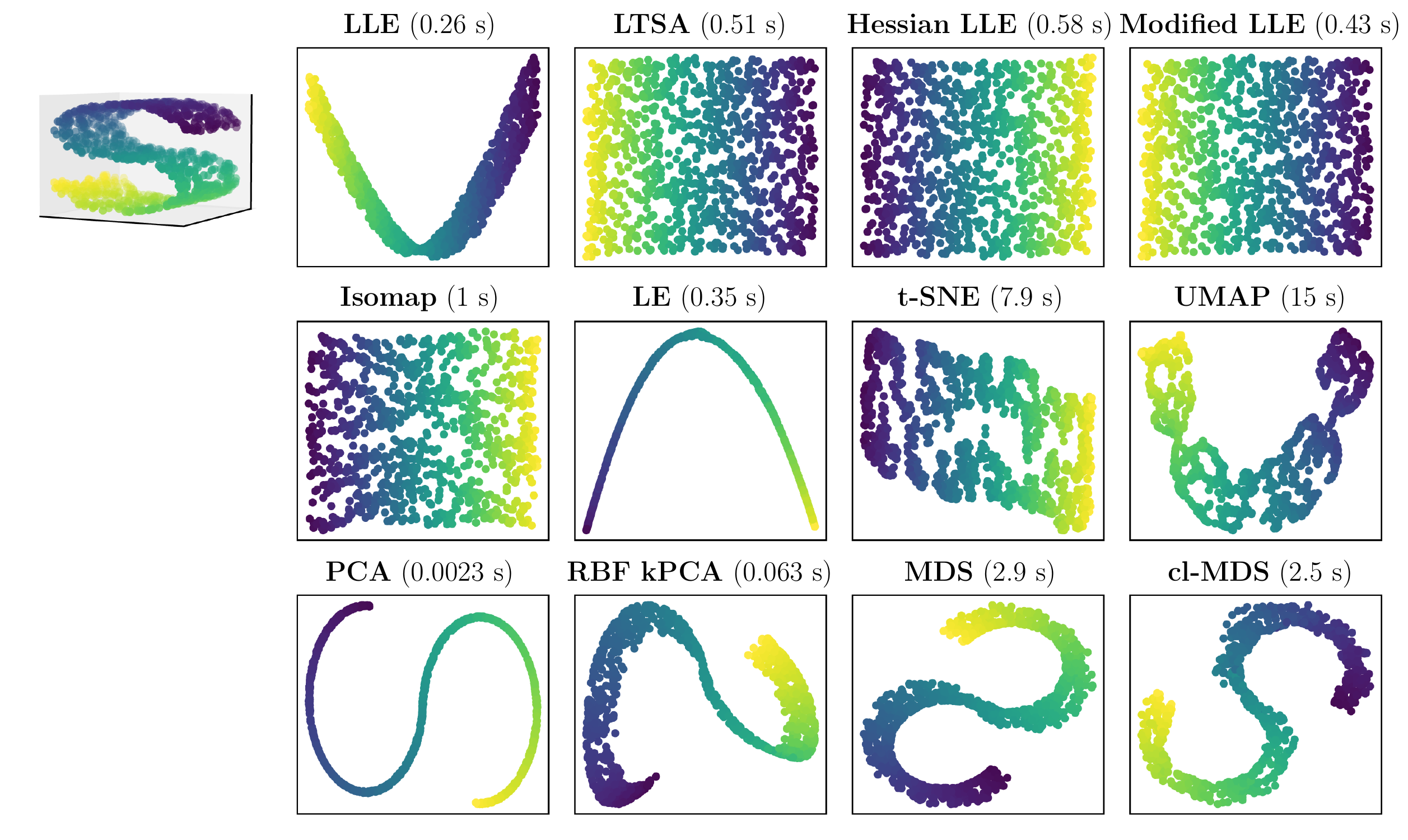}
    \caption{Comparison of several dimensionality reduction techniques applied to an S-curve manifold with 1000 points.
    Like the original example~\citep{S_example}, we used minimal parameters whenever possible. The top two rows include
    those methods that require a fixed number of neighbors, in this case limited to 15.}\label{fig:S_all_methods}
\end{figure}

On the other hand, the performance of cl-MDS improves speedwise with the number of clusters, being even
faster than MDS for $\ncl > 15$ (see MDS performance in Fig. \ref{fig:S_all_methods}). While counter-intuitive
at first, this result illustrates the usual trade-off between performance and sample size already mentioned in
\sect{ss_local}. Although the cl-MDS algorithm uses several instances of MDS, the subset of data points
processed per instance is much smaller than the complete sample. As Fig. \ref{fig:n_clusters} shows,
increasing $\ncl$ decreases average cluster size and speeds up individual cl-MDS calculations, until it
reaches a threshold 
($\ncl > 43$ in this example). This threshold is determined by the set of anchor points, whose size grows
approximately as $4\ncl$, recovering the previous tradeoff. Thus, we
conclude that our initial/previous compromise for choosing $\ncl$ is also reasonable in terms of performance.
In general, we expect that the computational complexity of cl-MDS scales similarly to those MDS instances, i.e., $\mathcal{O}(N^3)$ with $N$ switching between the total number of anchor points and the size of the biggest cluster depending on $\ncl$. However, the hierarchical embedding has a non-trivial effect in cl-MDS performance. On the other hand, preliminary empirical analysis for samples up to $N_\text{sample}=2000$ points showed an approximate scaling of $\mathcal{O}(N_\text{sample}^2)$. Further testing is left for future work, since we can characterize the computational complexity more appropriately once we have optimized the code thoroughly.

Note that, once a suitable $\ncl$ is reached, bigger clusterings may better incorporate the nuanced local details
and improve its visual quality, but they do not change the overall embedding significantly.
This is a fundamental distinction between cl-MDS and those methods whose
main hyperparameter fixes (or guesses) the \textit{number of close neighbors} for each point, i.e., the locality of the
embedding. Fig.~\ref{fig:S_all_methods} shows examples of the latter in its first two rows. Most of these methods aim to
embed the original data uniformly, as opposed to the techniques included in the last row which seek to preserve the
metric structure. That is, the former obtain an isotropic representation of the S-manifold but the metric information is
partially lost.

Additional cl-MDS parameters, such as the MDS weights and the percentile ranks for anchor points, were minimized here for
increased computational speed, reducing the accuracy too. In Fig. \ref{fig:n_clusters}, we can appreciate the extreme
linearity of certain cluster embeddings, which translates into the cl-MDS mapping being slightly less accurate than the
MDS one in Fig. \ref{fig:S_all_methods}. This effect is a consequence of the MDS optimization process, which does not
guarantee the preservation of relevant incidence relations (e.g., placing anchor points from the same cluster, originally
convex, on a line). Moreover, this is accentuated by a high number of clusters, poor choices of anchor points and
insufficient MDS iterations. To alleviate this issue, the cl-MDS implementation includes optimized values of all
related parameters by default, reducing the emergence of these artifacts.

\begin{figure}[t]
    \centering
    \includegraphics[width=0.8\columnwidth]{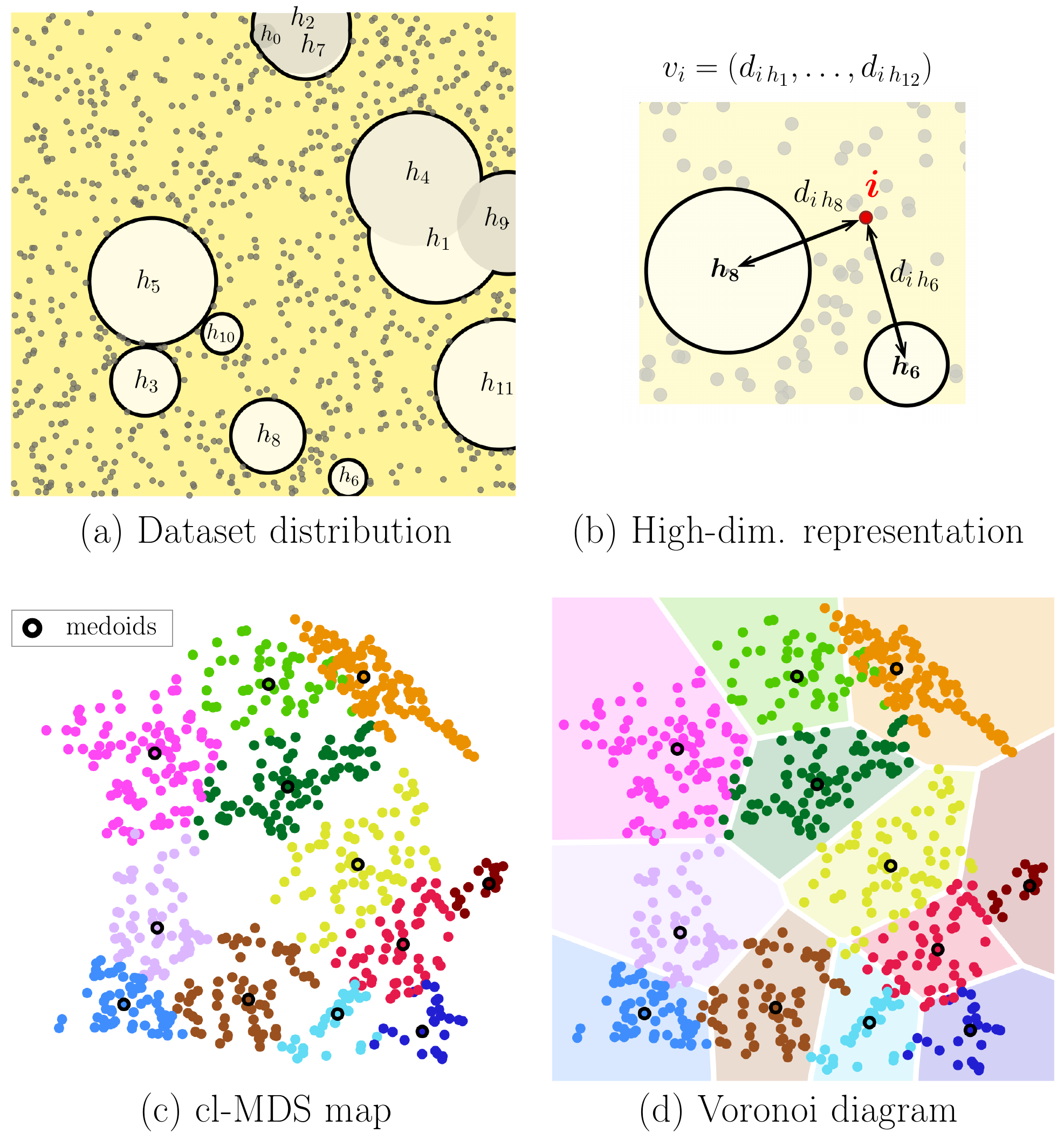}
    \caption{Simple example of cl-MDS applied to a high-dimensional dataset. The sample consists of 1000 points
    distributed in $\mathbb{R}^2$ as shown on panel (a), where $N_h = 12$. Their high-dimensional representation
    is obtained from the pairwise distances to each hole, illustrated on panel (b). Panels (c) and (d) show the
    cl-MDS embedding and the Voronoi diagram of the medoids, respectively.}\label{fig:swiss_cheese}
\end{figure}

The next example corresponds to a dataset of 1000 random points distributed over the unit square, avoiding $N_h$
randomly placed circular regions (\textit{holes}). Rather than characterizing each point $i$ by its 2-dimensional
coordinates, we use the vector $v_i:=(d_{i\,h_1},\hdots ,d_{i\,h_{N_h}})\in \mathbb{R}^{N_h}$ where $d_{i\,h_j}$
denotes its Euclidean distance to the center of the hole $h_j$, for $j=1,\hdots ,N_h$. This is a straightforward
approach for building a high-dimensional dataset with a custom dimension $N_h$. Figure~\ref{fig:swiss_cheese}
illustrates this for an $N_h=12$ example, with the pairwise distances in $\mathbb{R}^{12}$ fed to cl-MDS. Additionally,
this figure includes an application of Voronoi diagrams as a qualitative measure of accuracy for cl-MDS.
The Voronoi partition of a set of points corresponds to those regions of space, \textit{Voronoi cells},
containing the closest generator point. That is, the Voronoi partition associated to the medoids in $\mathbb{R}^{12}$
is equivalent, by definition, to clustering these medoids in this Euclidean space. Hence, the metric topology on this
partition is preserved in the 2-dimensional embedding only when the Voronoi cells in $\mathbb{R}^2$
contain the clusters perfectly. Figure~\ref{fig:swiss_cheese} shows how
close cl-MDS is to achieving this objective.

Now that we have built some intuition regarding the advantages, disadvantages and hyperparameters of cl-MDS, we can
apply it to more complex datasets. Also, advanced features such as sparsification and the corresponding estimation of
the complete 2-dimensional representation are used in the following examples.

\subsection{Visualizing atomic environments} \label{ss_atomic}
Analyzing atomic databases, which can comprise thousands (even millions) of structures, has become an
increasingly difficult task. During the last decade, new mathematical descriptions of atomic structures have
been developed as an alternative to simpler approaches, such as straightforward visualization of the structures,
which are prone to information overload from such databases. As a result, we can compare their members (either
atoms or molecules) using abstract mathematical representations known as \textit{atomic descriptors}
~\citep{bartok_soap, willatt_2019, de_2016}, where each member is characterized numerically. This representation
ranges from simple scalar distances and angles to complicated many-body atomic descriptors which are mapped
to high-dimensional vectors. These can be used to compute a similarity measure between atomic configurations based on kernel
functions, paving
the way to study these databases and their underlying properties. Still, the dimensionality of these descriptions
can obscure the results and their communication, as we discussed in \sect{s_intro}.

The high-dimensional nature of many-body atomic representations makes them a representative example of the
usefulness of
dimensionality reduction techniques, in general, and cl-MDS, in particular. These methods can be very
effective for visually comparing atomic structures and studying patterns within a database, fostering the
development of related software~\citep{de_2019_sketchmap,chemiscope,cheng_2020} in recent 
years. Some of the most common techniques adopted by materials scientists are the linear algorithm PCA,
its non-linear counterpart kernel PCA (kPCA)~\citep{kpca_1998}, metric MDS, 
sketch-map~\citep{ceriotti_2011}, and diffusion
maps~\citep{diff_maps_coifman_2005,diff_maps_coifman_2006}. While these methods are well established and extremely powerful
for visualization purposes, atomic-structure samples tend to be highly uneven (i.e., crucial data can 
be both over- and underrepresented) with multiple structural and chemical features at different scales. 
Even in the absence of noise, they can be challenging to embed and to display in a clear, intuitive way.

cl-MDS is designed to address those common pitfalls, especially for kernel similarities based on
smooth overlap of atomic positions (SOAP) descriptors~\citep{bartok_soap}. cl-MDS aims to enhance the visualization quality across several layers of information, i.e., dominant
features at different atomic length scales. On the other hand, atomic descriptors are increasingly used
as input for other ML models, either for analysis (e.g., classification tasks) or prediction purposes
(e.g., regression models for the potential energy). Adding their results on top of these visualizations is
straightforward, with the 2-dimensional embedding working as a ``white canvas''. We argue that the parallelism
between visual distances (i.e., Euclidean distances on a 2-dimensional plot) and kernel (dis)similarities
provides a transparent and intuitive baseline to navigate these complex samples.

All the examples in this section use the same framework. Following the specifics described
in Appendix \ref{app_atomic}, each data point corresponds to an atom represented
by a SOAP descriptor computed with the option \texttt{"quippy\_soap\_turbo"}. 
This way, the dataset has an associated kernel distance based on the similarities between atomic environments
defined within the chosen cutoff radii [see Eq.~(\ref{eq:kernel_dist})]. Note the
difference between using a cutoff sphere and restricting the number of neighbors as main criterion for
defining the idea of environment. The former establishes how far the neighborhood of an atom stretches
without assuming the number of nearest neighbors that populate that region. A cutoff sphere-based
representation allows us to systematically compare atoms with differing 
numbers of neighbors, more suitable for this application.

The scripts used to generate the data in the examples are  publicly available in the
cl-MDS GitHub repository~\citep{clmds}, except for the PtAu nanoclusters example. The latter is part
of a longer TurboGAP tutorial~\citep{turbogap}, where more visualizations are included. Other applications
of cl-MDS can be already found in the literature~\citep{vdw_muhli, Fe_nano_jana}.

\subsubsection{CHO structural database}\label{ss_cho}

\begin{figure}[t]
    \centering
    \includegraphics[width=1.\textwidth]{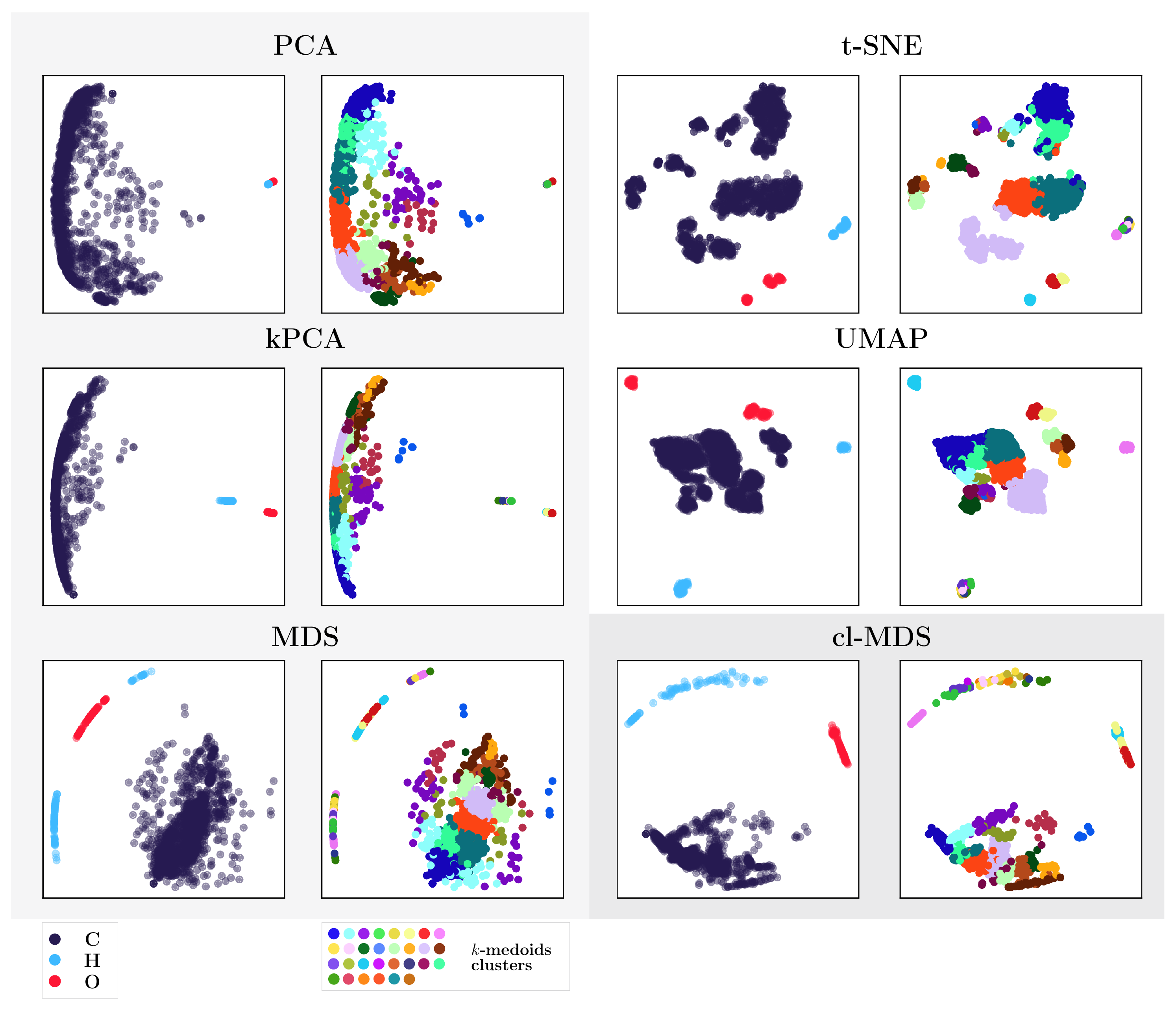}
    \caption{Embedding of 2000 randomly chosen data points performed using several methods, where each point corresponds to an
    atom from the CHO database. Two plots are included per method/embedding: one color-coded according to chemical species
    (C, H, O) and another color-coded according to $k$-medoids clustering.}
    \label{fig:CHO_comparison}
\end{figure}

The CHO database is a subset of the original dataset from Ref.~\citep{golze2021xps}, corresponding to a wide variety of
CHO-containing materials. It contains 675 computer-generated models of CHO materials, for a total
of 151,556 unique atomic environments.
We consider two approaches to dimensionality reduction: i) an overall embedding of a multispecies sample and ii) separated
embeddings per chemical species (C, H, O). The former is shown on Fig.~\ref{fig:CHO_comparison}, with several
dimensionality reduction techniques applied to a random sample of 2000 data points. Here, the high-dimensional representation was computed using cutoff radii 
$r_{\text{soft}}=3.75$~\AA, $r_{\text{hard}}=4.25$~\AA, with dimension $N_{\text{SOAP}}=2700$.
These radii ensured the convergence of the ML models trained in Ref.~\citep{golze2021xps} and, therefore,
retain enough relevant structural knowledge. Only those techniques that did not
fail were included, illustrating some of the common visualizations available for an atomic sample. The main reasons behind the poor visualizations obtained by Isomap, LE, LTSA, LLE and its variations are: (1) their assumption of a non-disconnected underlying manifold, which is very unlikely when different classes are present in a dataset (e.g., different atomic species); and (2) their tendency to collapse most of the data points together, due to poorly constrained cost functions. A more detailed analysis of the specific weaknesses of those methods can be found on Ref.~\citep{maaten_2008_dimred_review}.

First, it is noteworthy that the most elementary chemical intuition is captured visually by PCA, kPCA and cl-MDS
solely. A multispecies sample is characterized by higher dissimilarities among atoms from different chemical species,
unless the central atom is explicitly neglected. That is, we would expect that a 2-dimensional representation of such
sample reflects those dissimilarities with a proper separation of atomic species, particularly for metric-based
methods. However, all MDS attempts failed to encode global H and O dissimilarities in terms of the pairwise
distances in the embedded space, despite MDS being the quintessential method for pairwise distance preservation.
On the other hand, manifold-based methods such as t-SNE and UMAP are misleading in this regard, since neither
the distances nor the size of the clusters in their embeddings are meaningful. Even if they retain some global
structures, Fig.~\ref{fig:CHO_comparison} shows how relative positions between H clusters were lost.

Second, we incorporate cluster information from $k$-medoids into Fig.~\ref{fig:CHO_comparison}.  Given that data
clustering is independent of the embedding, we can compare how consistently each method preserves it. As expected,
cl-MDS outperforms other methods since its algorithm is based on clustering preservation, overcoming the tendency of
regular MDS to mix different clusters. 
Conversely, the usual ``clustered-appearance'' of t-SNE maps is only partially consistent with
$k$-medoids clustering. Meanwhile, PCA, kPCA and UMAP (to a lesser extent) tend to collapse smaller clusters irrespective of their
relevance, obscuring their visualization. In particular, this issue happened with UMAP despite increasing its
hyperparameter \verb|min_dist|, which adjusts the minimum distance between embedded points. Moreover, UMAP and t-SNE maps were obtained after performing hyperparameter estimation based on the silhouette score~\citep{rousseeuw_silhouette_score} for that very same clustering. We considered other two classifications, but they did not balance as well the local and global characteristics of the sample. The full details of these tests are included in the Supporting Information.

\begin{figure}[t]
    \centering
    \includegraphics[width=1\textwidth]{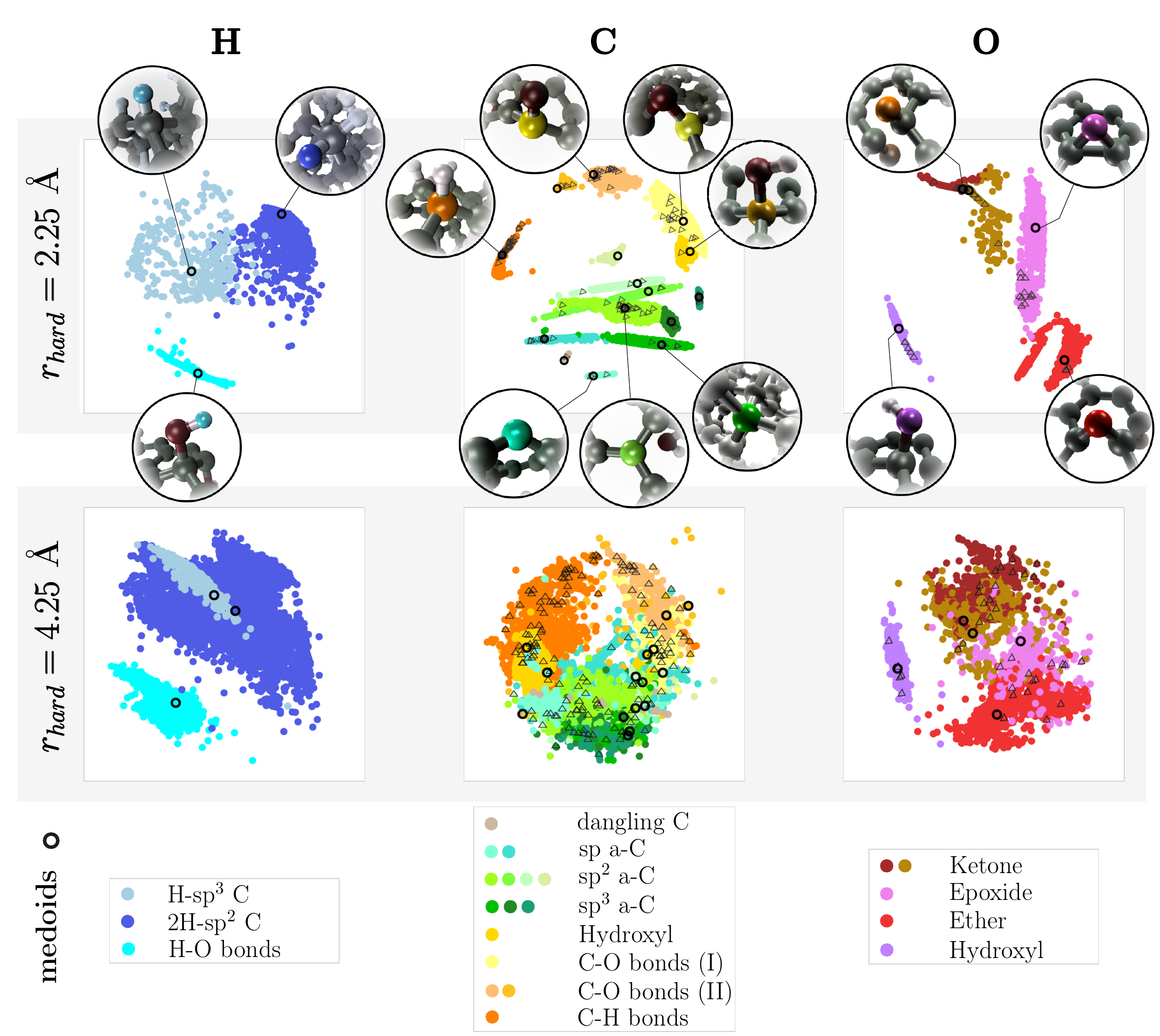}
    \caption{Visualization of the entire CHO database using cl-MDS. The embeddings were computed individually per
    chemical species (columns), considering high-dimensional representations with two different cutoff radii (rows).
    The medoids and clusters for the shortest radii ($r_{\text{soft}}=1.75$~\AA, $r_{\text{hard}}=2.25$~\AA) are
    highlighted in all embeddings, with their atomic motifs and their corresponding chemical denomination included.}
    \label{fig:CHO_final_map}
\end{figure}

Motivated by the clear separation between C, H and O atoms in Fig.~\ref{fig:CHO_comparison}, we computed the cl-MDS
maps per chemical species, as shown in Fig.~\ref{fig:CHO_final_map}. In this second approach, each embedding 
includes all atoms from the same species within the CHO database, i.e., 135,618 C atoms, 7669 H atoms and 8269 O
atoms. Sparsification support proved itself very handy,  especially for carbon, combined with the estimation of the
low-dimensional representation for the whole dataset (see Section~\ref{ss_hierarchy}). We found that a sparse set
containing as few as 2 percent of the whole carbon dataset was representative enough when carefully selected, e.g., using a
combination of random and CUR-based data points as well as precomputed medoids. Oxygen and hydrogen datasets, albeit
smaller in size, also benefited from sparsification due to its improvement of MDS performance within cl-MDS computations.

Since carbon has the richest structural landscape in this database, we eased the clustering and embedding process for
carbon by applying cluster hierarchy, with $h=[15,7,3,1]$ (see Section~\ref{ss_hierarchy}). We used a
simpler clustering for hydrogen and oxygen, with $N_{\text{cl}}=3$ and $N_{\text{cl}}=5$, respectively. Besides the
usual advantages of $k$-medoids versus $k$-means~\citep{arora_kmeans_vs_kmedoids} (e.g., allowing for different sizes
of clusters, being less sensitive to outliers), the knowledge of the medoids is itself valuable for visualization
purposes. That is, we do not only obtain a 2-dimensional plot of the dataset, but also a representative per cluster that
we can track back to the sample, as Fig.~\ref{fig:CHO_final_map} illustrates. We have added labels that identify the
clusters/medoids with classical chemical configurations (i.e., simple hybridizations and functional groups) to simplify
the visualization, associating several clusters with the same label. However, the clustering was thorough enough to
further distinguish between groups in terms of other structural features, such as angles or bond distances.

Additionally, Fig.~\ref{fig:CHO_final_map} contains two separate embeddings for each chemical species, computed from
SOAP representations of the database with different cutoff radii: the radii already used in Fig.~\ref{fig:CHO_comparison},
and a shorter version with $r_{\text{soft}}=1.75$~\AA, $r_{\text{hard}}=2.25$~\AA. These visualizations illustrate
the importance of hyperparameter selection depending on the ML application. While the SOAP representation does not change here its original
dimensionality (which depends exclusively on the number of basis
functions used), the region of SOAP space spanned by the sample increases, effectively becoming more sparsely
populated with increased averaged distances between data points, hindering data clustering.
Moreover, a small cutoff radius emphasizes first and second neighbors
in the representation, allowing an straightforward connection between the embeddings and classical chemical
motifs or functional groups in this example. 

On the other hand, large radii retain more structural information, despite increasing the
complexity of the atomic environments and their overall dissimilarities. That is, the larger cutoff radii
are better suited for training ML potentials and evaluating their performance, as opposed to the shorter
radii which ease data visualization and classification. For instance, the gray triangles in Fig.~\ref{fig:CHO_final_map}
indicate the sparse subset of atomic environments (for C and O atoms) used in Ref.~\citep{golze2021xps} to
build a ML model that involved expensive calculations. Their visualization in the first row proves that
their sampling process preserved the diversity of the database, since all motifs are represented. The sparse
set is not homogeneously distributed in those maps though, suggesting that a larger cutoff radii is needed for
the selection process, i.e., intuitive motif classification is not informative enough to train a predictive ML model. As
a final remark, note that the same radii could be too small for molecular datasets for
instance (see Section~\ref{ss_qm9}); in practice, the choice of radii depends heavily on the dataset and
the purpose of the visualization.

\subsubsection{QM9 database}\label{ss_qm9}

The QM9 database~\citep{qm9_2014} is a subset of a much larger database (the GDB-17 database~\citep{qm9_2012}, with 166
billion molecules) carefully selected for a detailed sampling of the chemical space of small organic compounds. In
particular, it contains 133,885 neutral organic molecules composed of carbon, hydrogen, oxygen, nitrogen and fluorine,
up to nine ``heavy'' atoms (C, O, N, F). To represent QM9 we use SOAP vectors of dimension $N_{\text{SOAP}}=7380$, and cutoff radii
$r_{\text{soft}}=3$~\AA, $r_{\text{hard}}=3.5$~\AA.
As discussed in Section~\ref{ss_cho}, the difference between chemical species outweighs any other dissimilarity in a
combined embedding; consequently, we performed a separate cl-MDS embedding per chemical species. A sparse set of
1000-2000 atoms depending on the species was used, carefully selected by combining $k$-medoids, random picking and a
consistent clustering. The visualization of each atomic species was obtained through the estimation of the
low-dimensional coordinates (see Section~\ref{ss_hierarchy}).

\begin{figure}[t]
    \centering
    \includegraphics[width=1\textwidth]{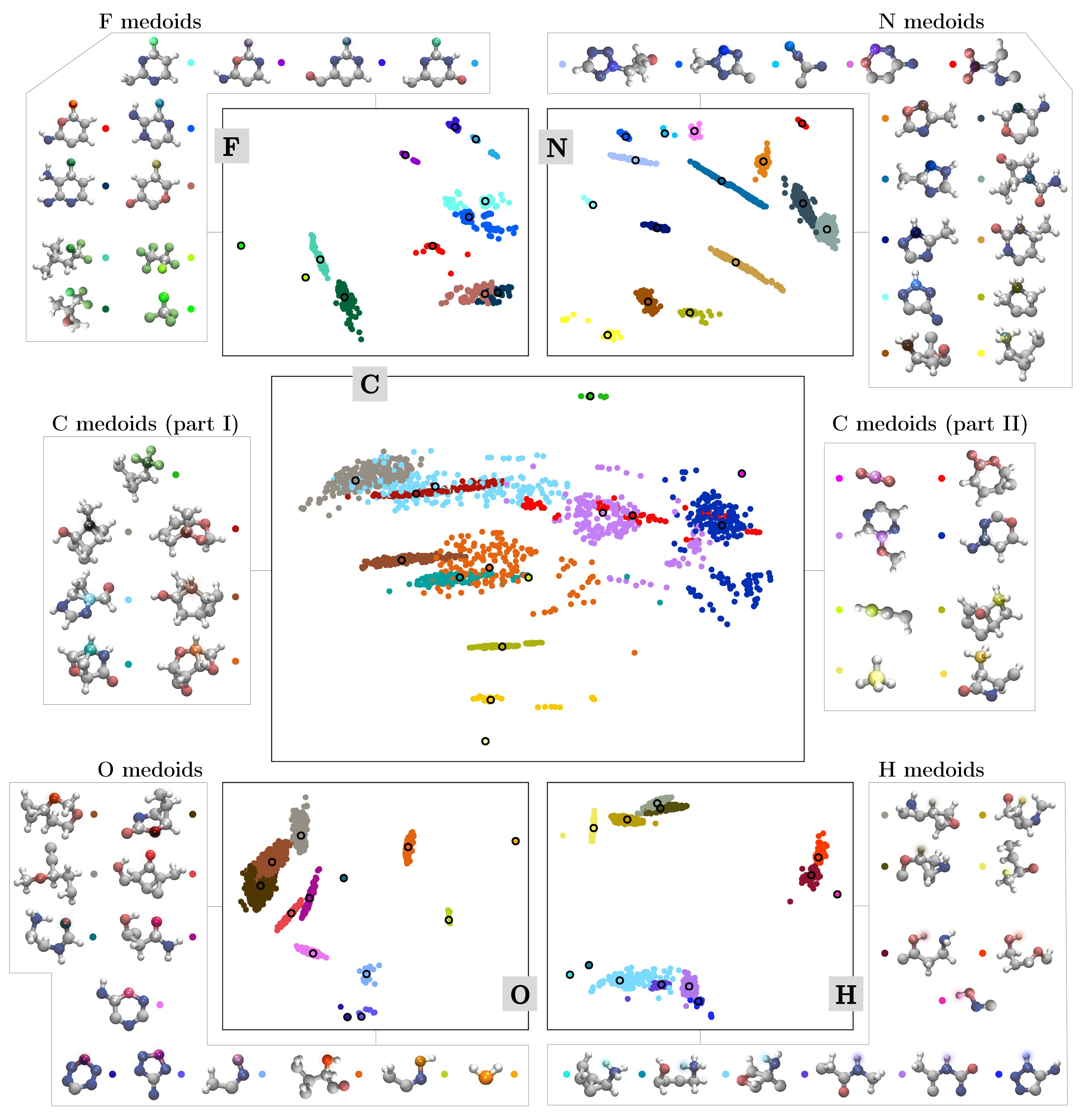}
    \caption{Visualization of the QM9 database through five cl-MDS mappings, one per atomic species (C, H, O, N, F).
             We show all the medoids and their atomic environments to illustrate the distribution of chemical and
             structural properties in each embedding, highlighting them in each molecular motif with the following
             colors: gray (C), white (H), red (O), blue (N) and green (F).}
    \label{fig:qm9}
\end{figure}

Figure~\ref{fig:qm9} shows the resulting cl-MDS embeddings, which help us visualize the composition of the QM9
database. Here we can appreciate once again how cl-MDS performs worse when separating clusters for chemical
species with richer variety of atomic environments, e.g., carbon. As opposed to the CHO example in Section~\ref{ss_cho} (see
Fig.~\ref{fig:CHO_final_map}), C atoms were embedded with a simple hierarchy, $h=[15,1]$, reducing the capability of
cl-MDS to effectively differentiate additional nuances. Its combination with medoids information nonetheless highlights
those subtleties via the corresponding molecular motifs, proving again the value of cluster information. Here, all the
medoids were included, revealing the existence of other relevant properties, apart from the chemical species, such as the
geometrical arrangement of the molecules. For instance, this is suggested by the absence of fluorine-related clusters
in H, N and O embeddings despite its presence in several structures. Independently of the number of clusters, fluorine
is not representative enough to weigh in the dissimilarities unless it is a first-nearest neighbor, i.e., with carbon. 

Additionally, we can easily identify those clusters whose members have extremely similar descriptors. While all clusters include at least 4 atomic configurations, the visualizations in Fig.~\ref{fig:qm9} show single-point clusters. This is a consequence of the high similarity between these atomic environments, which is reflected by their high-dimensional representation. That is, the kernel distances between those environments are effectively zero, representing them with the same point in both high- and low-dimensional representations.

\subsubsection{Database of PtAu nanoclusters} \label{ss_nps}

\begin{figure}[t]
    \centering
    \includegraphics[width=1\textwidth]{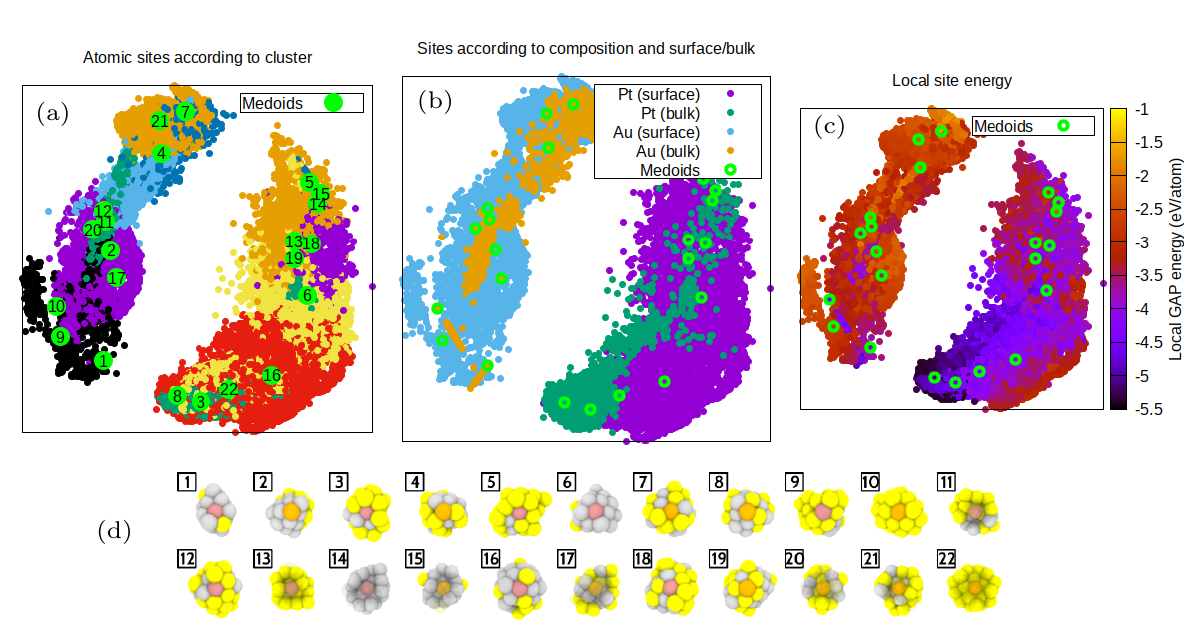}
    \caption{Visualization of a GAP-generated database of PtAu nanoclusters, using cl-MDS
    to illustrate (a) $k$-medoids clustering, (b) atomic sites in terms of composition and bulk/surface character,
    and (c) local energy predicted by the PtAu GAP. Those sites labeled as medoids are shown in panel (d),
    highlighting the central atom. Note that whenever a medoid in (d) in in the NP interior, some of the surface atoms are made transparent to allow for visualization of the medoid.}
    \label{fig:PtAu_clmds}
\end{figure}

In previous examples, we used visualizations to navigate the structural and chemical complexity of material and molecular databases. Now, we would like to introduce the potential of cl-MDS for the analysis and interpretation of scientific results obtained from other models, such as density-functional theory (DFT) or ML-driven atomistic simulations. For instance, we could assess the performance of a ML-based Gaussian approximation potential (GAP)~\citep{bartok_2010_gap} by generating a database and checking some physical and/or chemical properties on top of cl-MDS. We invite the reader to check the literature for further ideas about similar applications of dimensionality reduction methods, e.g., see Ref.~\citep{cheng_2020} for other examples in material science and chemistry.

For this example, we chose a general-purpose ML potential for modelling Pt and Au systems and computed a database of PtAu alloy nanoparticles (NPs). We want to better
understand the site distribution of this database, emphasizing any differences between ``bulk'' and
surface atoms, which might give useful insight in applications in catalysis. Note, however, that
there is no proper bulk in these NPs due to their reduced size. That said, the analysis workflow should be also applicable to larger structures without modification. The database was generated using the TurboGAP code~\citep{turbogap}, with 10
NPs per size and composition. It has 660 unique NPs, including pure Pt and Au ones. Each atomic site is
represented by a SOAP descriptor of size $N_\text{SOAP}=279$, using cutoff radii
$r_\text{soft}=5.2$ \AA, $r_\text{hard}=5.7$ \AA. In this case, we computed the representation using SOAP compression~\citep{klawohn_2023_compress}, which already reduces the dimension of the SOAP vector.

The resulting cl-MDS embedding is shown in Fig.~\ref{fig:PtAu_clmds}, with hierarchy $h=[22,1]$.
The entire set of atomic sites is 24750, with 500 sparse data points selected using \texttt{sparsify="cur"}.
When computing the full distance matrix is computationally affordable (memory-wise), CUR matrix decomposition is a suitable
sparsification option for evenly distributed samples, where relevant members are equally represented.
Those data points outside the sparse set were assigned estimated coordinates following Section~\ref{ss_hierarchy}
recipe. As expected, the clustering is influenced by the chemical species (Pt/Au) and its relative
position in the NP (surface/``bulk'').
Comparing Fig.~\ref{fig:PtAu_clmds} (b) and (c), we get a precise quantitative assessment of properties that we know to be true qualitatively, i.e., we can quantify and visualize chemical intuitive concepts. Here, the visualization illustrates how the cohesive energy of atoms naturally increases with atomic coordination, i.e., the latter produces a stabilizing effect moving from the interior sites to the surface sites.

\section{Conclusions and outlook}\label{s_conclusion}

We have introduced a novel technique for data visualization called cluster MDS, which aims to capture high-dimensional local and global features adequately in a single 2-dimensional representation. This issue is inadequately addressed by older methods due to limitations of their dimensionality reduction algorithms. More recent techniques still experience important limitations in this regard, such as the need for a ``balancing'' parameter that may crucially impact the structures preserved (e.g., UMAP, GLSPP~\citep{cai_GLSPP}), or the imposition of specific metrics that limit their application to other fields (e.g., PHATE, DGL~\citep{song_DGL}). 

The cl-MDS algorithm is based on a combination of data clustering and data embedding through $k$-medoids and metric MDS, respectively, with any distance matrix (or dissimilarity measure) as accepted input. We have illustrated the effect of its main hyperparameter, the number of clusters, which can capture the local nuances in the visualization without affecting the overall structure preservation, once a minimum value is reached. This value depends on the dataset and impacts the quality of cl-MDS mappings, rendering our heuristic selection in this paper unsatisfactory. We will explore available criteria to automate this choice and increase its reliability. In its more advanced form, this hyperparameter accepts a hierarchy of clusters that eases the embedding of highly complex data and aids the visualization of big amounts of unlabelled data.

Additionally, cl-MDS can estimate the 2-dimensional coordinates of other points once a first embedding has been obtained. Given that our method partially inherits the decreasing performance of MDS with dataset size, this estimation is extremely useful when combined with the included sparsification support. However, its quality strongly depends on the chosen sparse set and needs to be carefully assessed. We will update the options for automated sparse selection available in the code as more robust alternatives arise. Other future improvements of the code include the optimization of its computational speed and memory requirements.

Its comparison with well-known methods such as PCA, kPCA, t-SNE, MDS and UMAP showed that cl-MDS improved visualization of different layers of locality, most notably compared to MDS performance. We applied it to datasets of sizes $10^3$ to $10^6$ and dimensionality up to $7380$. In particular, we focused on atomic-structure examples to showcase all the functionality and advantages of this embedding tool, which includes specific recipes for atomic databases. Despite the value of manifold unfolding in other contexts, metric preservation is arguably the best approach for atomic databases. The comparison of atomic structures, as well as the study of their properties, usually involves some sort of similarity measure; therefore, its retention is invaluable in any visualization. Beyond the application to atomic structure datasets highlighted in this paper, we remark that cl-MDS is a general visualization tool. Any dataset with relevant local and global structures can benefit from its use, whenever the data accepts the definition of a meaningful metric. 

\acks{
The authors are grateful for financial support from the Academy of Finland under projects
\#321713, \#329483, \#330488 and \#347252, as well as computational resources provided by
CSC -- IT Center for Science and Aalto University's Science-IT Project.
}

\section*{Data Availability Statement}
The data that support the findings of this study are openly available: Figures 1--5 use data from Ref.~\citep{sklearn}, Figure~\ref{fig:qm9} uses QM9 data~\citep{qm9_2012, qm9_2014} and data from Figures 3 and 9 can be found in Ref.~\citep{clmds, turbogap} respectively.
The CHO database used in Figures 6 and 7 is available upon request from the authors from Ref.~\citep{golze2021xps}.

\appendix
\section{Embedding atomic structure representations}
\label{app_atomic}

We have added functionality to the cl-MDS code specifically tailored for processing atomic structure information.
First, the distance matrix $\mathbf{D}$ can be substituted by a file as input parameter, given in a format
compatible with the Atomic Simulation Environment (\verb|ASE|) Python library~\citep{ase_2002,ase_2017}.
This file may include a single atomic
structure, a concatenation of them, or a trajectory. Given this input, the user has three options: (1) choosing a cutoff radius, relying on the code defaults; (2) selecting a representation for the atomic descriptors,  being automatically computed by the cl-MDS code; or (3) providing an array of precomputed descriptors, e.g., obtained with an external tool such as \texttt{DScribe}~\citep{dscribe}.

Option (2) has currently three supported representations (\texttt{descriptor = "quippy\_soap" | "quippy\_soap\_turbo" | "quippy\_soap\_turbo\_compress"}) available via \verb|quippy|, a Python interface to
the Fortran code \verb|QUIP|~\citep{quip} generated by \verb|f90wrap|~\citep{f90wrap}. The first option, 
\texttt{"quippy\_soap"}, describes atomic environments using the smooth overlap of atomic positions (SOAP) vectors
~\citep{bartok_soap}. The  second option, \texttt{"quippy\_soap\_turbo"}, uses an optimized version~\citep{caro_2019} of the previous SOAP atomic descriptors through the \verb|soap_turbo| library~\citep{soap_turbo}. Additionally, a string of hyperparameter definitions (\texttt{descriptor\_string}) can be given for both descriptor options. The third option, \texttt{"quippy\_soap\_turbo\_compress"}, enables SOAP compression~\citep{klawohn_2023_compress} while using the default hyperparameters, computing similarly accurate SOAP vectors with lower dimensionality. Other implemented arguments include \texttt{average\_kernel},
which extends the previous mathematical representations to whole structures (only available with
\texttt{"quippy\_soap"}); and \texttt{do\_species}, which allows the pre-selection of certain species within a database to speed up the calculations.

Once we have a representation of the atomic structures, we need to obtain its associated distance matrix. A
suitably constructed atomic descriptor provides the basis for defining a metric that can be fed into the cl-MDS
algorithm. That is, these vectors are used to build a (usually) positive-definite kernel with an induced Gram
matrix $\mathbf{K}$ of size $N\times N$, where $N$ is the length of the database. 
For SOAP this kernel function is defined as~\citep{bartok_soap}
\begin{align}\label{eq:soap_kernel}
    K_{ij} = \left( \mathbf{q}^{\text{SOAP}}_{\,i}\cdot\, \mathbf{q}^{\text{SOAP}}_{\,j}\right)^\zeta\,,
\end{align}
where $\mathbf{q}^{\text{SOAP}}_{\,i}$ denotes the SOAP descriptor of atom $i$, $\zeta$ can be any positive scalar and $0\leq K_{ij}\leq 1$. The importance of defining a kernel resides
in its connection with a dissimilarity measure. As shown in Ref.~\citep{scholkopf_kernel_trick}, it is always
possible to find a dissimilarity measure $\mathbf{D}$ associated to a positive definite kernel $\mathbf{K}$,
given by
\begin{align}\label{eq:kernel_dist}
    (D_{ij})^2 = \frac{1}{2}(K_{ii}+K_{jj})-K_{ij}\,.
\end{align}
Considering polynomial kernels such as \eq{eq:soap_kernel} are always positive definite, the resulting SOAP dissimilarity matrix is
\begin{align}
    D_{ij} = \sqrt{1-K_{ij}}\,.
\end{align}
This distance measure is automatically computed by the cl-MDS code from the set of atomic descriptors prior
to performing the clustering and embedding steps. 

In addition to the embedding features described in \sect{ss_hierarchy}, we have also
added the possibility of fine-tuning the minimization process of the MDS stress from step \ref{step:4} (see
\sect{s_algorithm}) by means of a weighted distance matrix. While our MDS implementation can generally accommodate
weights, this is a straightforward extension for atomic structure visualization based on the SOAP kernel,
which is evaluated for the set of medoids
$\mathcal{M}$. That is, given two atoms $i \in \mathcal{C}_k$, $j\in  \mathcal{C}_s$ and a positive integer $\eta$,
the distance is redefined as
    \begin{align}
        D_{ij}^{(w)} = \sqrt{1 - K_{ij}\cdot (K_{m_k,m_s})^\eta }\,.
    \end{align}
    Thus, 
    \begin{align}
        \mathbf{D}^{(w)} = \left[ 1-\mathbf{K}\odot(\mathbf{K}_{\mathcal{M}})^{\odot \eta}\, \right]^{1/2}\,,
    \end{align}
        where $\odot$ denotes the element-wise product (also known as Hadamard or Schur product) and
        element-wise exponentiation, and
    \begin{align*}
        (K_{\mathcal{M}})_{\, ij} &= K_{m_k,m_s}\,,  &\text{if } i\in \mathcal{A}_k\text{ and } j\in  \mathcal{A}_s\,, \\
        (K_{\mathcal{M}})_{\, ij} &= 1, &\text{if and only if } i,j\in \mathcal{A}_k\,.
    \end{align*}
These weighted distances afford us greater control in decoupling the representation of individual clusters on the 
global map, effectively allowing us to continuously increase the emphasis on the local vs global structure of the data.

\end{document}